\documentclass[11pt,techReport]{eptcs}
\usepackage[utf8]{inputenc}
\usepackage{amsfonts}
\usepackage{amssymb}
\usepackage{amsmath}
\usepackage{amsthm}
\usepackage{units}
\usepackage{extarrows}
\usepackage{xspace,color}
\usepackage{enumitem}
\usepackage{multirow}

\usepackage{tikz,pgf}
\usetikzlibrary{trees,arrows,automata,positioning,plotmarks,matrix}

\usepackage{local}

\title{Breaking Symmetries}
\author{Kirstin Peters
\institute{Technische Universit\"at Berlin, Germany}
\email{kirstin.peters@tu-berlin.de}
\and
Uwe Nestmann
\institute{Technische Universit\"at Berlin, Germany}
\email{uwe.nestmann@tu-berlin.de}
}
\def\titlerunning{Breaking Symmetries}
\def\authorrunning{K.\ Peters, U.\ Nestmann}

\begin{document}
\maketitle

\begin{abstract}
A well-known result by Palamidessi tells us that \pimix (the $\pi$-calculus with mixed choice) is more expressive than \pisep (its subset with only separate choice).  The proof of this result argues with their different expressive power concerning leader election in symmetric networks.  Later on, Gorla offered an arguably simpler proof that, instead of leader election in symmetric networks, employed the reducibility of ``incestual'' processes (mixed choices that include both enabled senders and receivers for the same channel) when running two copies in parallel.  In both proofs, the role of \emph{breaking (initial) symmetries} is more or less apparent.  In this paper, we shed more light on this role by re-proving the above result|based on a proper formalization of what it means to break symmetries|without referring to another layer of the distinguishing problem domain of leader election.

Both Palamidessi and Gorla rephrased their results by stating that there is no uniform and reasonable encoding from \pimix into \pisep.  We indicate how the respective proofs can be adapted and exhibit the consequences of varying notions of uniformity and reasonableness.  In each case, the ability to break initial symmetries turns out to be essential.
\end{abstract}

\section{Introduction}

Palamidessi's well-known result~\cite{palamidessi03} tells us that \pimix (the $\pi$-calculus with mixed choice) is more expressive than \pisep (its subset with only separate choice).
More technically, the result states that there exists no ``good''---i.e., uniform (structure-preserving) and reasonable (semantics-preserving)---encoding from~\pimix into~\pisep.  Nestmann~\cite{nestmann00} proved that there is a "good" encoding from \pisep to \piasyn (the choice-free asynchronous subset of the $\pi$-calculus).  He also exhibited various encodings from  \pimix to \pisep, which were not considered ``good'' by Palamidessi, as they were not uniform or reasonable enough.
%That gives us some evidence to claim that \piasyn and \pisep have the same expressive power. Therefore the separation result of Palamidessi separates \pimix and \piasyn as well. 

Palamidessi's proof \cite{palamidessi03} argues with the different expressive power of the involved calculi concerning leader election in symmetric networks.  More precisely, Palamidessi proves that there is no symmetric network in \pisep that solves leader election, whereas there are such networks in \pimix. The proof implicitly uses the fact that it is not possible in \pisep to break initial symmetries, while this is possible in \pimix.  To this end, a rather strong notion of symmetry consisting of a syntactic and a semantic component is used to ensure that solving leader election requires breaking initial symmetries. With this result, inspired by Boug{\'e}'s work \cite{bouge88} in the context of \csp, Palamidessi proves that there is no uniform and reasonable encoding from \pimix into \pisep.

Later on, Gorla \cite{gorla08d} offered an arguably simpler proof for the non-existence of a ``good'' encoding from \pimix into \pisep. Instead of leader election in symmetric networks, it employed the reducibility of ``incestual'' processes (mixed choices that include both enabled senders and receivers for the same channel) when running two copies in parallel. Gorla's proof does not explicitly use a notion of symmetry.

Palamidessi's proof that there are no symmetric networks in \pisep that solve leader election addresses the \emph{absolute} expressive power of \pisep, whereas the proofs of the non-existence of a uniform encoding by Palamidessi and Gorla address the often-called \emph{relative} expressive power of the languages \cite{parrow08}. In the following, we discuss these two approaches in more detail, as this allows us to clarify the role of symmetry-breaking in the respective proofs.

The \emph{absolute expressive power} of a language describes what kind of behaviour or operations on behaviour are expressible in it (see \cite{parrow08,gorla08, gorla08d}). Analysing the absolute expressive power of a language usually consists of analysing which ``problems'' can be solved in it and which can not.  
It is often difficult to identify a suitable problem instance or problem domain to properly measure the expressive power of a language.  For instance, one might consider Turing-completeness to measure the computational power of a language.  In fact, Turing-completeness has been used in the context of process algebras, e.g., for Linda~\cite{busi00:_expres_of_linda_coord_primit}.   Instead, Palamidessi, inspired by Boug{\'e}~\cite{bouge88}, uses the distributed coordination problem of leader election.
%leading to rather strong Definitions of symmetry, 
More precisely, the problem refers to initially symmetric networks, where all potential leaders have equal chances and all processes run the same|read: symmetric|code.  There, to solve the leader election problem, it is required that in all possible executions a leader is elected.  Usually, it is argued that it is necessary|again in all possible executions|to break the initial symmetry in order to do so.  On the other hand, if there is just a single execution in which the symmetry is somehow perpetually maintained or at least restored, then also leader election may fail, and thus the leader election problem is not solved.
One may conclude that, at a closer look, Palamidessi's proof implicitly addresses another problem: the problem of breaking initial symmetries.  
Therefore, we suggest to promote ``breaking symmetries'' from a mere auxiliary proof technique to a proper problem of its own.  It turns out that, by doing so, we can significantly weaken the defintion of symmetry and at the same time provide a stronger proof applicable to problem instances different from leader election.

Now, to \emph{compare} the absolute expressive power of \emph{two} languages, we may simply choose a problem that can be solved in one language, but not in the other language.  
Actually, as soon as we compare two languages, it makes sense to use the term \emph{relative expressive power}, as we can now relate the two languages.  Unfortunately, the terminology was introduced differently.
It has been attributed (see~\cite{parrow08}) to the comparison of the expressive power of two languages by means of the existence or non-existence of encodings from one language into the other language, subject to various conditions on the encoding.%
\footnote{In our opinion, the denotation "relative expressive power" is misleading.  First, as mentioned above, also the absolute expressive power can directly be used to \emph{relate} two languages.  Second, %there is no common accepted notion of what conditions belongs to a "good" encoding|moreover there are hinds that such a common notion can not exists, because for different situations different sets of conditions turns out to be "good"|so the
results on the encodability of a language have to be understood relative to the specific conditions on the encoding---it is not always clear to what aspect the "relative" refers. Thus, in this paper, we prefer the notion of \emph{translational expressive power} to refer to comparisons of the expressiveness of two languages by analysing the existence or non-existence of an encoding, subject to various conditions.}
Both Palamidessi and Gorla state results of this kind; they prove that there is no uniform and reasonable encoding from \pimix into \pisep, for varying interpretations of the conditions uniform and reasonable.  

In this paper, we show that the problem of breaking inital symmetries, compared to the problem of leader election, appears to be a more suitable problem instance to separate \pimix from \pisep. There are two great benefits in proving an absolute separation result instead of a translational one. First, in opposite to translational separation results which are always equipped with the conditions on the encoding, we can formulate a separation result without any pre- or side conditions.  Second, as we show in Section \ref{sec:nonExistUniformEncoding}, we can prove several translational separation results due to different definitions of reasonableness %|including the settings used by Gorla in \cite{gorla08d}|
as simple consequences of our absolute separation result.
For our work, we had to develop answers to two related questions of definition:
\begin{itemize}
\item How exactly should one define \emph{symmetric} networks?
\item What exactly does it mean to \emph{break symmetries}?
\end{itemize}
\emph{The main contributions of this paper} are then as follows.
(1)~We present a separation result between \pimix and \pisep that does not require any additional preconditions.  In particular, it is completely independent of what it means for an encoding to be "good" or "reasonable". 
(2)~Since we use a weaker notion of symmetry, and because we do not focus on the leader election problem, our separation result is more general than the one in \cite{palamidessi03}, i.e., it widens the gap between \pimix and \pisep.  It also allows us to derive a number of translational separation results using counterexamples different from leader election.
(3)~We prove a stronger translational separation result in comparison to \cite{palamidessi03, vigliottiPhillipsPalamidessi07} and (the first setting of) \cite{gorla08d} by weakening the conditions on the encodings used.

\paragraph{Overview of the Paper.}

In \S\ref{sec:techn-prel}, we introduce the two process calculi that we intend to compare. In \S\ref{sec:SemanticVsSyntactic}, we revisit the notion of symmetry used by Palamidesi to propose her separation result and define symmetry as we  use it. In \S\ref{sec:breakingSymmetry}, we prove the separation result, i.e., we prove that \pimix is strictly more expressive as \pisep, by proving the inability of \pisep to break initial symmetries. Based on this result, we prove in \S\ref{sec:nonExistUniformEncoding} that there is no uniform and reasonable encoding from \pimix to \pisep examining different notions of reasonableness. We conclude with \S\ref{sec:conclusion}.

\section{Technical Preliminaries} \label{sec:techn-prel}

In the following, let $ \names $ denote a countable set of names. As is common nowadays, we present the $ \pi $-calculus including mixed guarded choice, but without match or mismatch operator \cite{sangiorgiWalker01, palamidessi03}.

\definition{def:fullPiCalculus}{$ \pi $-calculus}{
	The processes of the \textit{$ \pi $-calculus}, denoted by $ \procmix $, are given by
	\begin{align*}
		P \quad & ::= \quad \sum_{i}{\alpha_i.P_i} \sep P \mid P \sep \newVar{z}{P} \sep \; !P \quad \text{, where}\\
		\alpha \quad & ::= \quad \outU{x}{y} \sep x\left( z \right) \sep \tau
	\end{align*}
}
\noindent
Note that the process term $ \sum_{i}{\alpha_i.P_i} $ represents \emph{finite} guarded choice; as usual, the term ${\alpha_1.P_1}+{\alpha_2.P_2}$ denotes binary choice,  and we use $ \nullTerm $ as abbrevation for the empty sum.

%In asynchronous communication the sending and receiving of messages is not instantaneously but rather time consuming. Hence a sending process can not base its further progress on the successful receiving of the sent message by an other process. To implement these restriction in the asynchronous $ \pi $-calculus an output can be used only as guard for $ \nullTerm $.
%\definition{def:asynPiCalculus}{Asynchronous $ \pi $-calculus}{
%	The processes of the asynchronous \textit{$ \pi $-calculus}, denoted by \piasyn, are given by
%	\begin{align*}
%		P \quad ::= \quad \outU{x}{\tilde{y}}.\nullTerm \sep \sum_i{\alpha_i.P_i} \sep P \mid P \sep \newVar{z}{P} \sep \; !P
%	\end{align*}
%	where
%	\begin{align*}
%		\alpha \quad ::= \quad x\left( \tilde{z} \right) \sep \tau
%	\end{align*}
%}
In the $ \pi $-calculus with separate choice, both output and input can be used as guards, but within a choice term either there are no input or no output guards, i.e., we have input and output guarded choice, but no mixed choice.
\definition{def:PiCalculusWithSeparateChoice}{$ \pi $-calculus with separate choice}{
	The processes of the \textit{$ \pi $-calculus with separate choice}, denoted by \procsep, are given by
	\begin{align*}
		P \quad & ::= \quad \sum_i{\alpha_i^I.P_i} \sep \sum_i{\alpha_i^O.P_i} \sep P \mid P \sep \newVar{z}{P} \sep \; !P \quad \text{, where}\\
		\alpha^I \quad & ::= \quad x\left( z \right) \sep \tau \hspace*{3em} \text{and} \hspace*{3em}
		\alpha^O \quad ::= \quad \outU{x}{y} \sep \tau
	\end{align*}
}
We use $ x, x', x_1, \ldots, y, y', y_1, \ldots, z, z', z_1, \ldots $ to range over names and capital letters $ P, P', P_1, \ldots, Q, R, \ldots $ to range over processes. We often omit $ \nullTerm $ in longer terms. If we refer to processes without further requirements we mean elements of \procmix; we sometimes use just $\proc$ when the discussion applies to both.

Let $ \act \deff \Set{ x{y},\; \outU{x}{{y}},\; \outB{x}{{y}} \mid x,y \in \names } $ denote the set of visible actions, where $x{y}$ denotes \emph{free input}, $ \outU{x}{y} $ denotes \emph{free output} and $ \outB{x}{{y}} $ denotes \emph{bound output}. Let $ \tau $ denote an internal not visible action. Let $ \lab $ be the corresponding set of \emph{labels}, i.e., $ \lab = \act \cup \Set[]{ \tau } $. We use $ \mu, \mu', \mu_1, \ldots $ to range over labels. Let $ \freenames{P} $ and $ \freenames{\mu} $  denote the sets of \emph{free names} in $ P $ and $\mu$, respectively. Let $ \boundnames{P} $ and $ \boundnames{\mu} $ denote the sets of \emph{bound names} in $P$ and $ \mu $, respectively. Likewise, $ \allnames{P} $  and $ \allnames{\mu} $ denote the sets of all \emph{names} occurring in $P$ and $ \mu$. Their definitions are completely standard. We assume that there are no clashes between free and bound names in terms, i.e., in any term the set of bound and free names are disjoint.% We explicitly allow that names are bound several times within a term.

The operational semantics of \procmix and \procsep are jointly given by the transition rules in Figure \ref{fig:operationalSemantics}, where congruence $ \equiv $ is defined (according to \cite{palamidessi03}) by the following rules:
\begin{enumerate}
	\item $ P \; \equiv \; Q $ if $ Q $ can be obtained from $ P $ by alpha-conversion
	\item $ \newVar{x}\!{P} \mid Q \; \equiv \; \newVar{x}{\left(  P \mid Q  \right)} $ if $ x \notin \freenames{Q} $
	\item $ P \mid Q \; \equiv \; Q \mid P $
\end{enumerate}
%\U{Welche Definition von $\equiv$ stellst Du Dir vor?  Die von [Pal03]? (Lies dazu auch mal ihre Bemerkungen auf S10 von MSCS.)  Das ist wichtig zu wissen, denn das beeinflusst die Wahrheit der Aussagen \"uber die wieder-erreichten Netzwerke.  Wenn die syntaktisch nur durch ein $\sigma$ unterschieden werden sollen, dann darf man mit $\alpha$-conversion nciht zu freiz\"ugig umgehen; oder man definiert ``Gleichheit'' prinzipiell immer modulo $\alpha$-Konversion. Catuscia's Regeln aus [Pal03] erlauben beliebige gebundene Umbenennungen vor und nach Schritten, und auch die Scope Extrusion in der Richtung, die wir nach unseren heutigen Skypen wohl auch brauchen.  Ich denke, die sollten wir vorerst nehmen.  Ich denke, wir brauchen in $\textsc{Cong}$ ``vorher'' $\alpha$ und ``nachher'' extrusion.  [Pal03] diskutiert das \"ubrigens auch auf S14 von MSCS.}
\begin{figure}[ht]
	\begin{align*}
		\begin{array}{|c|}
			\hline
			\\
			\textsc{I-Sum} \quad \sum_i{\alpha_i.P_i} \step{xy} \Set{ \Subs{y}{z} } P_j \quad a_j = x\left( z \right) \hspace*{3em} \textsc{O/$ \tau $-Sum} \quad \sum_i{\alpha_i.P_i} \step{\alpha_j} P_j \quad \alpha_j = \outU{x}{y} \text{ or } \alpha_j = \tau\\
			\\
			\textsc{Par} \quad \dfrac{P \step{\mu} P'}{P | Q \step{\mu} P' | Q} \quad \boundnames{\mu} \cap \freenames{Q} = \emptyset\\ %\hspace*{3em} \textsc{Par-r} \quad \dfrac{P \step{\mu} P'}{Q | P \step{\mu} Q | P'} \quad \boundnames{\mu} \cap \freenames{Q} = \emptyset\\
			\\
			\textsc{Comm} \quad \dfrac{P \step{\outU{x}{y}} P' \quad Q \step{xy} Q'}{P | Q \step{\tau} P' | Q'} \hspace*{3em} %\textsc{Comm-R} \quad \dfrac{Q \step{\out{x}y} Q' \quad P \step{xy} P'}{P | Q \step{\tau} P' | Q'}\\
%			\\
			\textsc{Close} \quad \dfrac{P \step{xy} P' \quad Q \step{\outB{x}{y}} Q'}{P \mid Q \step{\tau} \newVar{y}{\left(P' \mid Q'\right)}} \quad y \notin \freenames{P}\\ %\hspace*{2em} \textsc{Close-R} \quad \dfrac {P \step{\outB{x}{y}} P' \quad  Q \step{xy} Q'}{P \mid Q \step{\tau} \newVar{y}{\left(P' \mid Q'\right)}} \quad y \notin \freenames{Q}\\
			\\
			\textsc{Res} \quad \dfrac{P \step{\mu} P'}{\newVar{z}{P} \step{\mu} \newVar{z}{P'}} \quad z \notin \allnames{\mu} \hspace*{3em} \textsc{Rep} \quad \dfrac{P \mid !P \step{\mu} P'}{!P \step{\mu} P'}\\
			\\
			\textsc{Open} \quad \dfrac{P \step{\outU{x}{y}} P'}{\newVar{y}{P \step{\outB{x}{y}} P'}} \quad x \neq y \hspace*{3em} \textsc{Cong} \quad \dfrac{P' \equiv P \quad P \step{\mu} Q \quad Q \equiv Q'}{P' \step{\mu} Q'}\\
			\\
            %\text{The symmetric versions \textsc{Par-R}, \textsc{Comm-R}, \textsc{Close-R} of \textsc{Par-L}, \textsc{Comm-L}, \textsc{Close-L} are omitted.}
            %\\
			\hline
		\end{array}
	\end{align*}
	\caption{Operational semantics} \label{fig:operationalSemantics}
\end{figure}

As usual, the tuple notation $ \tilde{x} \in \tupel{\names} $ denotes finite sequences $x_1,\ldots,x_n$ of names in $\names$, i.e., $ \tupel{M} $ denotes the set of tuples over a set $ M $. Moreover, we use $ \newVar{\tilde{x}}{} $ for a sequence $ \tilde{x} = x_1, \ldots, x_n $ to abbrevate $ \newVar{x_1}{\ldots \newVar{x_n}{}} $ and $ \tilde{x} \setminus M $ for a set of names $ M $ to denote the sequence of names $ \tilde{x} $ without the occurrences of name $ y $ for all $ y \in M $. We also use the tuple notation for other kinds of data, like actions or labels.

A \emph{network} is a process $ \newVar{\tilde{x}}{\left( P_1 \mid \ldots \mid P_n \right)} $ for some $ n \in \Nat $, $ P_1, \ldots, P_n \in \proc $ and $ \tilde{x} \in \tupel{\names} $. We refer to $ P_1, \ldots, P_n $ as the processes of the network.

We use $ \sigma $, $ \sigma' $, $ \sigma_1 $, \ldots to range over substitutions. A substitution is a set $ \Set{ \Subs{x_1}{y_1}, \ldots, \Subs{x_n}{y_n} } $ of rules to rename free names of a term. $ \Set{ \Subs{x_1}{y_1}, \ldots, \Subs{x_n}{y_n} }\left( P \right) $ is defined as the result of replacing all occurences of $ y_i $ by $ x_i $ for $ i \in \Set{ 1, \ldots, n } $, possibly applying alpha-conversion to avoid capture or name clashes. For all names $ \names \setminus \Set{ y_1, \ldots, y_n } $ the substitution behaves as identity function. Let $ \id $ denote identity, i.e., $ \id $ is the empty substitution $ \id = \Set[]{ } $.% We often omit the brackets, i.e., $ \sigma\left( P \right) = \sigma P $.

As usual, $ P \step{\mu} P' $ denotes a step from $ P $ to $ P' $, where $ \mu $ is either a label of an action or $ \tau $. Moreover let $ P \step{} $ ($ P \nStep{} $) denote existence (non-existence) of a step from $ P $, i.e., there is (no) $ P' \in \proc $ and (no) $ \mu \in \lab $ such that $ P \step{\mu} P' $. A (partial) execution is a sequence of steps $ P \step{\mu_1, \ldots, \mu_n} P' $ such that $ P \step{\mu_1} H_1 \step{\mu_2} \ldots \step{\mu_{n{-}1}} H_{n-1} \step{\mu_n} P' $ for some $ P', H_1, \ldots, H_{n{-}1} \in \proc $ with the sequence $ \mu_1, \ldots, \mu_n $ of observable and unobservable actions, i.e., $ \mu_1, \ldots, \mu_n \in \lab $. Accordingly $ P \step{\tilde{\mu}} P' \nStep{} $ denotes a finite execution from $ P $ to $ P' $ with the sequence of actions $ \tilde{\mu} \in \tupel{\lab} $.% \K{Besser? Ich hab mal nachgeschlagen. Trace wird in Walker/Sangiorgi ohne $ \tau $s definiert. Ich brauche die $ \tau $s aber auch, also spreche ich wohl lieber von Sequenzen von Aktionen, auch wenn das etwas länger ist.}\U{Eventuell ginge ``strong traces'' \dots und dann als Konvention, dass wir, da wir in diesem Papier ohnehin keine ``weak traces'' verwenden, einfach ``traces'' sagen.}

\section{Semantic versus Syntactic Symmetry} \label{sec:SemanticVsSyntactic}

Palamidessi in \cite{palamidessi03} proved that \pimix is strictly more expressive than \pisep by proving that the former can solve leader election in symmetric networks while the latter can not. The leader election problem consists of choosing a leader among the processes of a network. In \cite{palamidessi03}, a special channel $ \mathit{out} $ is assumed to propagate the index of the winning process, i.e., the leader. The leader election problem is solved by a network iff in each of its executions each process propagates the same process index over $ \mathit{out} $ and no other index is propagated.

As already Boug\'{e} did for \csp in \cite{bouge88}, Palamidessi uses a \emph{semantic} definition of symmetry.  Intuitively, the \emph{syntactic component} of the symmetry definition in \cite{bouge88, palamidessi03, vigliottiPhillipsPalamidessi07} states two processes as symmetric iff they are identical modulo some renaming according to a permutation $ \sigma $ on their free names. Boug\'{e} \cite{bouge88} argues why a syntactic notion of symmetry does not suffice considering the leader election problem to distinguish \cspmix, i.e., \csp where input and output commands may appear in guards, and \cspin, i.e., \csp where only input commands may appear in guards. He presents two networks in \cspin each solving leader election although each should be considered as syntactically symmetric. %\footnote{Note, that using our definition of symmetry none of these networks solves leader election. Intuitively this is because we allow no variables in terms and so the symmetry relation handles the instantiated $ i $ and $ 0 $ in the first example and the instantiations of $ name $ and $ i $ in the second example respectively in exactly the same way.}\U{Eine Fussnote, die auf Beispiele in einem anderen Papier verweist? Lieber nicht.  War die am Freitag schon da?}. 
The following example presents such a syntactically symmetric network solving leader election in \pisep:
\begin{align}
	N \; \triangleq \; P \mid \sigma\left( P \right) \quad \text{ with } \quad P = \out{x} \mid x.\outU{\mathit{out}}{\; 1} + y.\outU{\mathit{out}}{\; 2} \quad \text{ and } \quad \sigma = \Set{ \nicefrac{x}{y}, \nicefrac{y}{x} } \label{exm:symSolLE}
\end{align}
$ N $ is syntactically symmetric with respect to the permutation $ \sigma $, i.e., $ N = P_1 \mid P_2 $ and $ P_2 $ is equal to $ P_1 $ modulo the exchange of $ x $ and $ y $ according to $ \sigma $. Moreover $ N $ solves the leader election problem.

To overcome these problems the \emph{semantic component} of the symmetry definition is designed to be strongly connected to the problem considered, i.e., leader election in this case. Intuitively, its purpose is to ensure that the only way to solve the leader election problem is to break the initial symmetry of the given network. Note that $ N $ does \emph{not} break the initial syntactic symmetry, because e.g. in the execution $ N \step{\tau} P \mid \outU{\mathit{out}}{\; 1} \step{\tau} \outU{\mathit{out}}{\; 1} \mid \outU{\mathit{out}}{\; 1} \step{\outU{\mathit{out}}{\; 1}} \nullTerm \mid \outU{\mathit{out}}{\; 1} \step{\outU{\mathit{out}}{\; 1}} \nullTerm \mid \nullTerm \not\step{} $ each second step results in a network that is syntactically symmetric with respect to $ \sigma $.  So, without this semantic part in the definition of symmetry, the leader election problem can not be used to distinguish \pimix and \pisep (or \cspmix and \cspin).%: there are syntactic symmetric networks in \pisep that solve leader election, e.g., the network $ N $ in (\ref{exm:symSolLE}) above.  

\paragraph{Semantic symmetry.}
\label{sec:semantic-symmetry}

We revisit Palamidessi's notion of symmetry for the $ \pi $-calculus as of \cite{palamidessi03}. Note that the involved definitions are based on the ones introduced by Boug\'{e} in \cite{bouge88} for \csp.

According to \cite{palamidessi03}, a hypergraph is a tuple $ H = \left\langle N, X, t \right\rangle $, where $ N $ and $ X $ are finite sets whose elements are called nodes and edges and $ t $, called type, is a function assigning to each edge the set of nodes connected by this edge. An automorphism on a hypergraph is a pair $ \sigma = \left\langle \sigma_N, \sigma_X \right\rangle $ such that $ \sigma_N: N \to N $ and $ \sigma_X: X \to X $ are permutations which preserve the type of edges. Given a hypergraph $ H $ and $ \sigma $ on $ H $ the orbit of a name $ n $ is the set of nodes in which the iterations of $ \sigma $ map $ n $.

A network $ P \equiv \newVar{\tilde{x}}{\left( P_1 \mid \ldots \mid P_k \right)} $ of $ k $ processes solves the leader election problem if for every computation of $ P $ there exists an extension of the computation and there exists an index $ n \in \Set{ 1, \ldots, k } $ such that for each process the extended computation contains one output action of the form $ \outU{\mathit{out}}{\; n} $ and no other action $ \outU{\mathit{out}}{\;m} $ with $ m \neq n $. The hypergraph associated to a network $ P $ is the hypergraph $ H(P) = \left\langle N, X, t \right\rangle $ with $ N = \Set{ 1, \ldots, k } $, $ X = \freenames{P_1 \mid \ldots \mid P_k} \setminus \Set{ \mathit{out} } $, and for each $ x \in X $, $ t(x) = \Set{ n \mid x \in \freenames{P_n} } $. Given a network $ P $ and the hypergraph $ H(P) $ associated to $ P $ an automorphism on $ P $ is any automorphism $ \sigma = \left\langle \sigma_N, \sigma_X \right\rangle $ on $ H(P) $ such that $ \sigma_X $ coincides with $ \sigma_N $ on $ N \cap X $ and $ \sigma_X $ preserves the distinction between free and bound names.

A network $ P $ with the associated hypergraph $ H(P) = \left\langle N, X, t \right\rangle $ and an automorphism $ \sigma $ on $ P $ is symmetric with respect to $ \sigma $ iff for each node $ i \in N $, $ P_{\sigma(i)} \equiv_{\alpha} \sigma\left( P_i \right) $\footnote{In \cite{bouge88} and \cite{vigliottiPhillipsPalamidessi07} formally slightly different conditions but with the same effect are used.}, holds where $ \equiv_{\alpha} $ denotes equality modulo alpha conversion.

To distinguish \pimix and \pisep Palamidessi shows that a network $ P \in \procsep $ which is symmetric with respect to an automorphism $ \sigma $ on $ P $ with only one orbit can not solve the leader election problem while this is possible in \pimix.

The main point of the semantic component of symmetry is that the special channel $ \mathit{out} $ can not be renamed by $ \sigma $ while the indices of the processes of the network must be permuted by $ \sigma $.  With that, the network $ N $ in (\ref{exm:symSolLE}) above is \emph{not} symmetric according to \cite{palamidessi03}.  This allows Palamidessi to prove that for each execution of a network in \procsep, which is symmetric with respect to an automorphism $ \sigma $, whenever there is an output $ \outU{\mathit{out}}{\; i} $ there is an output $ \outU{\mathit{out}}{\; \sigma\left( i \right)} $ with $ \sigma\left( i \right) \neq i $ as well, which contradicts the leader election problem. This explains why in \cite{bouge88, palamidessi03, vigliottiPhillipsPalamidessi07} such an effort is spent to define symmetry.

Nevertheless it turns out that we do not need the leader election problem to distinguish \pimix and \pisep. The main argument in the proof of \cite{palamidessi03} that there is no symmetric network in \procsep solving leader election is that it is impossible in \pisep to break symmetries.

\paragraph{Syntactic symmetry.}

As mentioned in the introduction, we directly focus on the problem of breaking symmetries instead of concentrating on leader election. Thus, we can release most of the above conditions for symmetry. Moreover, we abandon the notion of hypergraphs and automorphisms. Instead, we use a simple syntactic definition of symmetry that, as mentioned above, states two processes as symmetric iff they are identical modulo some renaming according to a permutation $ \sigma $ on their free names.

\definition{def:symmetryRelation}{Symmetry relation}{
	A \textit{symmetry relation of degree $ n $} is a permutation $ \sigma : \names \to \names $, such that $ \sigma^n = \id $.
	
	Let $ \symRels{n}{\names} $ denote the set of symmetry relations of degree $ n $ over $ \names $ and let $ \sigma^0 = \id $.
	
	% We say that a name $ x \in \names $ is in the scope of $ \sigma $ if and only if $ \sigma\left( x \right) \neq x $.
}
\noindent
Note that this definition does not require that $ n $ is the minimal degree of $ \sigma $; consequently, the condition that $ \sigma $ is an automorphism with only one orbit is released. %  Moreover, it strengthens the statement of Theorem \ref{thm:cannotBreakSymmetry} allowing the subdivition of a symmetric execution.
A symmetric network is then a network of $ n $ processes that are equal except for some renaming according to a symmetry relation $ \sigma $.

\definition{def:symmetricNetwork}{Symmetric network}{
  Let $ P \in \proc $. 
  Let sequence $ \tilde{x} $ contain only free names of $ P $.
  Let $ n \in \Nat $.
  Let $ \sigma $ be a symmetry relation of degree $ n $ over $ \names \setminus \boundnames{P} $.
  Let $\tilde{x}$ be closed under $\sigma$.
  Then
  \begin{align*}
    \SymNetwork{P}{\sigma}{n,\tilde{x}} = \newVar{\tilde{x}}{\big(\;\symNetwork{P}{\sigma}{n}\;\big)}
  \end{align*}
  is a \textit{symmetric network of degree $ n $}.
}
\noindent
Note that, in the following proofs, we make use of the fact that names bound in $ P $ are bound in each other process of $ \SymNetwork{P}{\sigma}{n,\tilde{x}} $ as well, so we explicitly forbid alpha-conversion here. In the following, whenever we assume some symmetric network $\SymNetwork{P}{\sigma}{n,\tilde{x}}$, we implicitly assume the respectively quantified parameters: a process $P\in\proc$, a sequence $\tilde{x}$ containing only free names of $P$, a network size $n\in\Nat$, a symmetry relation $\sigma$ of degree $ n $ over $ \names \setminus \boundnames{P} $.

The main difference of our definition to the definition of a symmetric network in \cite{palamidessi03} is that, in \cite{palamidessi03}, the processes of a symmetric network are numbered consecutively and for each process $ P_i $ within the symmetric network $ P_{\sigma\left( i \right)} \equiv \sigma\left( P_i \right) $ holds.  Thus, each symmetric network in \cite{palamidessi03} is a symmetric network for our definition, but not vice versa. Our definition of symmetry is weaker.

%\U{Palamidessi hat \"ubrigens eine All-Quantifikation bzgl.\ der Automorphismen in ihrer Definition; wir haben nur Existenz.  Was bedeutet das?}\K{Ich hab keine Ahnung, warum Palamidessi das mit rein genommen hat. Sie benutzt das selbst nicht.}

We use an index-guided form of substitution to replace single processes within a symmetric network.

\definition{def:indexedSubstitution}{Indexed substitution}{%\U{Ich w\"urde sie eventuell auch nur im Rahmen des Beweises als ``inlined def'' definieren. ``For convenience".  Aber vielleicht betont es ganz gut den Vorteil der Definition von symmetrischen Netzwerken, wenn man es schon hier tut. Du kannst/solltest(?) au{\ss}erdem auch explizit die Analogie zu Vektoren ziehen, die unsere symmetrischen Netzwerke ja faktisch sind.}
	Let $ \SymNetwork{P}{\sigma}{n, \tilde{x}} $ be a symmetric network. An \textit{indexed substitution} of some processes within a symmetric network, denoted by $ \Set{ \HOSubs{Q_1}{i_1}, \ldots, \HOSubs{Q_m}{i_m} }\SymNetwork{P}{\sigma}{n, \tilde{x}} $ for some processes $ Q_1, \ldots, Q_m \in \proc $ and $ i_1, \ldots, i_m \in \Set{ 0, \ldots, n{-}1 } $ such that for all $ j, k \in \Set{ 1, \ldots, m } $ $ j \neq k $ implies $ i_j \neq i_k $, is the result of exchanging $ \sigma^{i_k}\left( P \right) $ in $ \SymNetwork{P}{\sigma}{n, \tilde{x}} $ by $ Q_k $ for all $ k \in \Set{ 1, \ldots, m } $.
}
\noindent
Obviously $ \Set{ \HOSubs{Q_1}{i_1}, \ldots, \HOSubs{Q_m}{i_m} }\SymNetwork{P}{\sigma}{n, \tilde{x}} $ is a network; in general, however, it is not symmetric with respect to $ \sigma $.

\section{Symmetric Executions} \label{sec:breakingSymmetry}

We explicitly prove that in \pisep it is not possible to break initial symmetries, i.e., starting with a symmetric network there is always at least one execution preserving the symmetry. We  refer to such a execution as \emph{symmetric execution}. Let us consider a symmetric network $ \SymNetwork{P}{\sigma}{n, \tilde{x}} $ of degree $ n $. %:
%\begin{align*}
%	\SymNetwork{P}{\sigma}{n, \tilde{x}}
%\end{align*}
Of course, if only one process does a step on its own, then all the other processes of the network can mimic this step and thus restore symmetry. So, there is a symmetry preserving execution if there is no communication between the processes of the network. The most interesting case is how the symmetry is restored after a communication between two processes of the network has temporarily destroyed it. Both cases are reflected in the proof of Theorem \ref{thm:cannotBreakSymmetry}.

Apart from symmetric networks, we use the notion of a symmetric sequence of actions. Similarly to symmetric networks, in which a symmetry relation is applied to processes to derive symmetric processes, a symmetric sequence of actions is the result of applying a symmetry relation to action labels. It is sometimes necessary to translate a bound output to an according unbound output because a network can send a bound name several times but only the first of this outputs will be bound.

\definition{def:symmetricBehavior}{Symmetric sequence of actions}{
	Let $ \mu \in \lab $ be an action label, let $ \tilde{x} \in \tupel{\names} $ be a sequence of names and $ \sigma $ a symmetry relation of degree $ n \in \Nat $. Then $ \SymSequence{\mu}{\sigma}{n, \tilde{x}} $ denotes the sequence $ \mu_1, \ldots, \mu_n $ of $ n $ labels such that $ \mu_1, \ldots, \mu_n \in \lab $, $ \mu_1 = \mu $ and for $ i \in \Set{ 2, \ldots, n } $:
	\begin{align*}
		\mu_i = \begin{cases} \tau, & \text{ if } \mu = \tau\\ \sigma^i\left( a \right)b, & \text{ if } \mu = ab\\ \outU{\sigma^i\left( a \right)}{\sigma^i\left( b \right)}, & \text{ if } \mu = \outU{a}{b} \OR \left( \mu = \outB{a}{b} \AND \sigma^i\left( b \right) \notin \tilde{x} \setminus \Set{ b, \sigma\left( b \right), \ldots, \sigma^{i{-}1}\left( b \right) } \right)\\ \outB{\sigma^i\left( a \right)}{\sigma^i\left( b \right)}, & \text{ if } \mu = \outB{a}{b} \AND \sigma\left( b \right) \in \tilde{x} \setminus \Set{ b, \sigma\left( b \right), \ldots, \sigma^{i{-}1}\left( b \right) } \end{cases}
	\end{align*}
	
	Sometimes we refer to $ \mu_2, \ldots, \mu_n $ as the symmetric counterparts of $ \mu $.
}

Intuitively, a symmetric execution is an execution starting from a symmetric network returning to a symmetric network after any $ n $'th step, and which is either infinite or terminates in a symmetric network. Thereby, each sequence of $ n $ steps is labeled by a symmetric sequence of actions. 
%Intuitively, a symmetric execution is a execution starting from a symmetric network returning to a symmetric network after any $ n $'th step, and which is either infinite or terminates in a symmetric network; it also has a special kind of observable behavior, denoted as symmetric behavior. More precisely, in the $ n {-} 1 $ steps used to restore the symmetry destroyed by the first step the processes of the network mimic the first step with respect to the symmetry relation and with it they mimic its observable behavior.

\definition{def:symmetricTrace}{Symmetric execution}{
	A \textit{symmetric execution} is either a finite execution of length $ m \in \Nat $
	\begin{align*}
		\SymNetwork{P}{\sigma}{n, \tilde{x}} \step{\SymSequence{\mu_1}{\sigma_1}{n, \tilde{x}}} \SymNetwork{P_1}{\sigma_1}{n, \tilde{x}_1} \step{\SymSequence{\mu_2}{\sigma_2}{n, \tilde{x}_1}} \ldots \step{\SymSequence{\mu_m}{\sigma_m}{n, \tilde{x}_{m{-}1}}} \SymNetwork{P_m}{\sigma_m}{n, \tilde{x}_m} \not\step{}
	\end{align*}
	for some $ P_1, \ldots, P_m \in \proc $, $ \mu_1, \ldots, \mu_m \in \lab $, $ \tilde{x}_1, \ldots, \tilde{x}_m \in \tupel{\names} $ and $ \sigma_1, \ldots, \sigma_m \in \symRels{n}{\names} $ such that $ \sigma \subseteq \sigma_1 \subseteq \ldots \subseteq \sigma_m $ or an infinite execution
	\begin{align*}
		\SymNetwork{P}{\sigma}{n, \tilde{x}} \step{\SymSequence{\mu_1}{\sigma_1}{n, \tilde{x}}} \SymNetwork{P_1}{\sigma_1}{n, \tilde{x}_1} \step{\SymSequence{\mu_2}{\sigma_2}{n, \tilde{x}_1}} \SymNetwork{P_2}{\sigma_2}{n, \tilde{x}_2} \step{\SymSequence{\mu_3}{\sigma_3}{n, \tilde{x}_2}} \ldots \text{.}
	\end{align*}
	for some $ P_1, P_2, \ldots \in \proc $, $ \mu_1, \mu_2, \ldots \in \lab $, $ \tilde{x}_1, \tilde{x}_2, \ldots \in \tupel{\names} $ and $ \sigma_1, \sigma_2, \ldots \in \symRels{n}{\names} $ such that $ \sigma \subseteq \sigma_1 \subseteq \sigma_2 \subseteq \ldots $.
}
\noindent
Note that because of $ \sigma \subseteq \sigma_1 \subseteq \ldots $ the symmetry relation can only increase during a symmetric execution such that existing symmetries are preserved. Moreover|as shown in Lemma \ref{lem:cannotBreakSymmetry}|the symmetry relation does only grow in the presence of bound output to capture the renaming done by alpha-conversion. In the absence of bound output we have $ \sigma = \sigma_1 = \ldots = \sigma_m $ and $ \sigma = \sigma_1 = \sigma_2 = \ldots $ respectively.

Palamidessi proved that \pisep enjoys a certain kind of \textit{confluence} property \cite{palamidessi03}. Let $ \outP{x}{y} $ denote an output action, i.e., $ \outP{x}{y} $ is either a bound output $ \outB{x}{y} $ or an unbound output $ \outU{x}{y} $.
\lemma{lem:localConfluence}{
	Let $ P \in \procsep $ be a process. If $ P $ can make two steps $ P \step{\outP{x}{y}} Q $ and $ P \step{zw} R $ then there exists $ S $ such that $ Q \step{zw} S $ and $ R \step{\outP{x}{y}} S $.
}
\begin{proof}[Proof of Lemma \ref{lem:localConfluence}]
	See proof of Lemma 4.1 in \cite{palamidessi03} at pages 17 to 18.
\end{proof}

With this property we prove that it is not possible to break symmetries in \pisep.  Intuitively, we show that there is at least one symmetric execution by proving that whenever there is a step destroying symmetry we can restore it in $ n {-} 1 $ more steps mimicking the first step. The respective existence relies on the standard Lemma in process calculi like the $\pi$-calculus that transitions are preserved under substitution.  As conclusion it is not possible in \pisep to break an initial symmetry in all executions.

\theorem{Symmetric Execution}{thm:cannotBreakSymmetry}{
  Every symmetric network in \procsep has at least one symmetric execution.
}
\begin{proof}[Proof of Theorem \ref{thm:cannotBreakSymmetry}]
	We first prove the following statement:
	\lemma{lem:cannotBreakSymmetry}{
		\begin{align*}
			\begin{array}{c}
				\forall n \in \Nat \logdot \forall \tilde{x} \in \tupel{\names} \logdot \forall P \in \procsep \logdot \forall \sigma \in \symRels{n}{\names \setminus \boundnames{P}} \logdot \forall \mu \in \lab \logdot\\
				\SymNetwork{P}{\sigma}{n, \tilde{x}} \step{\mu} \widehat{P} \IMP \exists P' \in \procsep \logdot \exists \tilde{x}' \in \tupel{\names} \logdot \exists \mu_2, \ldots, \mu_n \in \lab \logdot \exists \sigma' \in \symRels{n}{\names} \logdot\\
				\widehat{P} \step{\mu_2, \ldots, \mu_n} \SymNetwork{P'}{\sigma'}{n, \tilde{x}'} \AND \mu, \mu_2, \ldots, \mu_n = \SymSequence{\mu}{\sigma'}{n, \tilde{x}} \AND \sigma \subseteq \sigma'
			\end{array}
		\end{align*}
	}
	Intuitively it states that given an arbitrary symmetric network $ \SymNetwork{P}{\sigma}{n, \tilde{x}} $ in \procsep, whenever $ \SymNetwork{P}{\sigma}{n, \tilde{x}} $ can perform a step then there are exactly $ n {-} 1 $ more steps that restore symmetry, i.e., that lead to a symmetric network again and the corresponding $ n $ steps are labeled by a sequence of symmetric actions. Note that the main line of argumentation of this Lemma is very similar to the proof of Theorem 4.2 in \cite{palamidessi03} at pages 18 to 23, although we prove a completly different statement. Nevertheless due to the different proof statements the proofs differ in technical details.% We only present an informal proof outline here. A more formal proof can be found in the appendix.

\begin{proof}[Proof of Lemma \ref{lem:cannotBreakSymmetry}]
	$ \SymNetwork{P}{\sigma}{n, \tilde{x}} \step{\mu} \widehat{P} $ can be the result of either an internal $ \mu $-step of one proccess of the network, i.e., it can be produced without the rules \textsc{Comm} or \textsc{Close}, or of a communication between two processes of the network, i.e., be produced by one of the rules \textsc{Comm} or \textsc{Close}. In the first case only one process performs a step and the rest of the network remains equal, i.e.:
	\begin{align*}
		\exists i \in \Set{ 0, \ldots, n{-}1 } \logdot \exists H \in \procsep \logdot \exists \tilde{x}_1 \in \tupel{\names} \logdot \sigma^i\left( P \right) \step{\mu} H \AND \widehat{P} \equiv \Set{ \HOSubs{H}{i} }\SymNetwork{P}{\sigma}{n, \tilde{x}_1} \tag{C1} \label{caseC1}
	\end{align*}
	In the second case $ \mu = \tau $ and two processes of the network change, i.e.:
	\begin{align*}
		\begin{array}{c}
			\exists i, j \in \Set{ 0, \ldots, n{-}1 } \logdot \exists H_1, H_2 \in \procsep \logdot \exists z, z' \in \names \logdot i \neq j \AND \Big( \sigma^i\left( P \right) \mid \sigma^j\left( P \right) \step{\tau} H_1 \mid H_2\\
			\OR \sigma^i\left( P \right) \mid \sigma^j\left( P \right) \step{\tau} \newVar{z, z'}{\left( H_1 \mid H_2 \right)} \Big) \AND \widehat{P} \equiv \Set{ \HOSubs{H_1}{i}, \HOSubs{H_2}{j} }\SymNetwork{P}{\sigma'}{n, \tilde{x}'}
		\end{array} \tag{C2} \label{caseC2}
	\end{align*}
	We proceed with a case split.
	\begin{description}
		\item[Case (\ref{caseC1}):] Within the symmetric network $ \SymNetwork{P}{\sigma}{n, \tilde{x}} $ all processes $ \sigma^i\left( P \right) $ for $ i \in \Set{ 0, \ldots, n{-}1 } $ are equal except for some renaming of free names according to $ \sigma $. That means whenever a process $ \sigma^i\left( P \right) $ can perform a step $ \step{\mu} $ then each other process $ \sigma^k\left( P \right) $ of the network can mimic this step by $ \step{\mu'} $, where $ \mu' $ is the result of applying $ \sigma^{k{-}i{+}n} $ to $ \mu $ possibly by changing bound output to unbound output as described in Definition \ref{def:symmetricBehavior}.\footnote{Note that $ n $ is added in $ k{-}i{+}n $ and $ k{-}j{+}n $ just to ensure that both values are positive. Because $ \sigma^n = \id $ if $ k{-}i \geq 0 $ we have $ \sigma^{k{-}i{+}n} = \sigma^{k{-}i} $.} The case of a bound ouput action $ \mu $ is a little bit tricky so we consider the other cases first. If $ \mu $ is no bound output we can choose $ \mu_2, \ldots, \mu_n $ such that $ \mu, \mu_2, \ldots, \mu_n = \SymSequence{\mu}{\sigma}{n, \tilde{x}} $. Moreover by symmetry $ \sigma^i\left( P \right) \step{\mu} H $ implies $ \sigma^k\left( P \right) \step{\mu'} \sigma^{k -i{+}n}\left( H \right) $ for a $ H \in \procsep $. With it we can restore symmetry by mimicking the $ \mu $-step of process $ \sigma^i\left( P \right) $ by the $ n {-} 1 $ steps $ \sigma^{i{+}1}\left( P \right) \step{\mu_2} \sigma\left( H \right) $, \ldots, $ \sigma^{n{-}1}\left( P \right) \step{\mu_{n{-}i}} \sigma^{n{-}1{-}i}\left( H \right) $, $ \sigma^0\left( P \right) \step{\mu_{n{-}i{+}1}} \sigma^{n{-}i}\left( H \right) $, \ldots, $ \sigma^{i{-}1}\left( P \right) \step{\mu_n} \sigma^{n{-}1}\left( H \right) $. These $ n $ steps build the chain
			\begin{align*}
				\begin{array}{lcl}
					\SymNetwork{P}{\sigma}{n, \tilde{x}} & \step{\mu} & \newVar{\tilde{x}_1}{\left( \sigma^0\left( P \right) \mid \ldots \mid \sigma^{i{-}1}\left( P \right) \mid \sigma^0\left( H \right) \mid \sigma^{i{+}1}\left( P \right) \mid \ldots \mid \sigma^{n{-}1}\left( P \right) \right)}\\
					& \vdots\\
					& \step{\mu_{n{-}i}} & \newVar{\tilde{x}_{n{-}i}}{\left( \sigma^0\left( P \right) \mid \ldots \mid \sigma^{i{-}1}\left( P \right) \mid \sigma^0\left( H \right) \mid \ldots \mid \sigma^{n{-}1{-}i}\left( H \right) \right)}\\
					& \step{\mu_{n{-}i{+}1}} & \newVar{\tilde{x}_{n{-}i{+}1}}{\left( \sigma^{n{-}i}\left( H \right) \mid \sigma\left( P \right) \mid \ldots \mid \sigma^{i{-}1}\left( P \right) \mid \sigma^0\left( H \right) \mid \ldots \mid \sigma^{n{-}1{-}i}\left( H \right) \right)}\\
					& \vdots\\
					& \step{\mu_n} & \newVar{\tilde{x}_n}{\left( \sigma^{n{-}i}\left( H \right) \mid \ldots \mid \sigma^{n{-}1}\left( H \right) \mid \sigma^0\left( H \right) \mid \ldots \mid \sigma^{n{-}1{-}i}\left( H \right) \right)}
				\end{array}
			\end{align*}
			with $ \tilde{x}_1, \ldots, \tilde{x}_n \in \tupel{\names} $ and $ \tilde{x}' = \tilde{x}_n $. Because $ \sigma^n = \id $ after the last step we result in a network which is again symmetric with respect to $ \sigma $, i.e., we choose $ \sigma' = \sigma $. With that we can choose $ P' = \sigma^{n{-}i}\left( H \right) $ such that $ \SymNetwork{P}{\sigma}{n, \tilde{x}} \step{\mu} \widehat{P} \step{\mu_2, \ldots, \mu_n} \SymNetwork{P'}{\sigma'}{n, \tilde{x'}} $. If $ \mu $ is an input or an unbound output action then so are its symmetric counterparts $ \mu_2, \ldots, \mu_n $, so we can choose $ \tilde{x}' = \tilde{x}_1 = \ldots = \tilde{x}_n = \tilde{x} $ and are done. Else if $ \mu $ is a bound output action $ \outB{y}{z} $ we have to consider two cases.
			\begin{description}
				\item[Case $ z \notin \boundnames{\sigma^i\left( P \right)} $:] Then of course $ z \in \freenames{\sigma^i\left( P \right)} $ and because $ \mu $ is a bound output $ z $ must be in $ \tilde{x} $. So we have to choose $ \tilde{x}_1 = \tilde{x} \setminus \Set{ z } $. Then by Definition \ref{def:symmetricBehavior} some of the actions $ \mu_2, \ldots, \mu_n $ might be bound and some might be unbound outputs depending on whether $ \sigma^{j{-}1}\left( z \right) $ was already the subject of an earlier bound output of this sequence or not. If $ \sigma^{j{-}1}\left( z \right) $ of $ \mu_j $ was already the subject of an bound output within $ \mu, \mu_2, \ldots, \mu_{j{-}1} $ then $ \mu_j $ is an unbound output and with it we choose $ \tilde{x}_j = \tilde{x}_{j{-}1} $, else $ \mu_j $ is a bound output and we choose $ \tilde{x}_j = \tilde{x}_{j{-}1} \setminus \sigma^{j{-}1}\left( z \right) $ for all $ j \in \Set{ 2, \ldots, n } $. Again we can choose $ \sigma' = \sigma $ and $ P' = \sigma^{n{-}i}\left( H \right) $ and proceed as in the case where $ \mu $ is no bound output.
				\item[Case $ z \in \boundnames{\sigma^i\left( P \right)} $:] Then by symmetry $ \sigma^j\left( z \right) = z $ is bound in $ \sigma^{i{+}j}\left( P \right) $ for all $ j \in \Set{ 0, \ldots, n{-}1 } $. By the above assumption that there are no name clashes we conclude $ z \notin \tilde{x} $. Then by Definition~\ref{def:symmetricBehavior} $ \mu, \mu_2, \ldots, \mu_n $ is a sequence of $ n $ bound output actions. Each of these actions $ \mu_j $ changes the scope of $ \sigma^{i{+}j{-}1}\left( P \right) $ (in a symmetric way to the other processes) but the scope of the network is left unchanged. With it we can choose $ \tilde{x}' = \tilde{x}_1 = \ldots = \tilde{x}_n = \tilde{x} $ again. The crux is that performing the first bound output with label $ \mu $ may force an alpha-conversion to avoid name clashes to the other bound instances of $ z $ in the other processes of the network such that the symmetry would be destroyed. To illustrate this problem let us consider an example first:
				\begin{description}
					\item[Example:] Let $ N \; \triangleq \; \newVar{x}{\outU{a}{x}.\out{x}} \mid \newVar{x}{\outU{a}{x}.\out{x}} = \SymNetwork{\newVar{x}{\outU{a}{x}.\out{x}}}{\id}{2} $. $ N $ can perform two bound outputs $ \outB{a}{x} $. To avoid name capture we have to apply alpha-conversion such that we have $ N \step{\outU{a}{x}} \out{x} \mid \newVar{x'}{\outU{a}{x'}.\out{x'}} \step{\outU{a}{x'}} \out{x} \mid \out{x'} $. Because of this alpha-conversion we result in a network which is not symmetric with respect to $ \id $. To overcome this problem we record the renaming done by alpha conversion in $ \sigma' = \Set{ \Subs{x}{x'}, \Subs{x'}{x} } $ such that $ \out{x} \mid \out{x'} = \SymNetwork{\out{x}}{\sigma'}{2} $. Note that because of this $ \sigma' $ can only increase by adding permutations on formerly bound names and fresh names.
				\end{description}
				That is why we have to increase the symmetry relation in this case to keep track of the renaming done by alpha-conversion. Thereto we enforce that the alpha-conversion after the first bound output to rename all instances of $ z $ (except to first) to a different fresh name for each process of the network and add the respective permutations of $ z $ to $ \sigma $ in order to obtain $ \sigma' $ such that $ \sigma \subseteq \sigma' $. Afterwards we can choose $ \mu_2, \ldots, \mu_n $ such that $ \mu, \mu_2, \ldots, \mu_n = \SymSequence{\mu}{\sigma'}{n, \tilde{x}} $ and $ P' = \sigma'^{n{-}i}\left( H \right) $ and proceed as in the case where $ \mu $ is no bound output.
			\end{description}
		\item[Case (\ref{caseC2}):] In this case there is a communication between $ \sigma^i\left( P \right) $ and $ \sigma^j\left( P \right) $ as result of one of the rules \textsc{Comm} or \textsc{Close}. Without loss of generality let us assume that $ \sigma^i\left( P \right) $ is the sender and $ \sigma^j\left( P \right) $ is the receiver of this communication, i.e., there are $ y, z_1, z_2 \in~\names $ such that $ \sigma^i\left( P \right) \step{\outP{y}{z_1}} H_1 $ and $ \sigma^j\left( P \right) \step{y z_2} \Set{ \Subs{z_2}{z_1} }\left( H_2 \right) $. Because of symmetry each process $ \sigma^k\left( P \right) $ for $ 0 \leq k \leq n{-}1 $ can perform an output action $ \mu_{\mathsf{out}, k} = \outP{\sigma^{k{-}i{+}n}\left( y \right)}{\sigma^{k{-}i{+}n}\left( z_1 \right)} $ (either bound or unbound) and an input action $ \mu_{\mathsf{in}, k} = \sigma^{k{-}j{+}n}\left( y \right) z_2 $ such that $ \sigma^k\left( P \right) \step{\mu_{\mathsf{out}, k}} \sigma^{k{-}i{+}n}\left( H_1 \right) $ and $ \sigma^k\left( P \right) \step{\mu_{\mathsf{in}, k}} \sigma^{k{-}j{+}n}\left( \Set{ \Subs{z_2}{z_1} }\left( H_2 \right) \right) $. Because of local confluence (compare to Lemma \ref{lem:localConfluence}), i.e., without mixed-choice an output action can not block an alternatively input action (within one step) and vice versa\footnote{Indeed without mixed choice the only possibility for $ \sigma^k\left( P \right) $ to be able to perform both actions is that these two actions are composed in parallel, so $ \sigma^k\left( P \right) $ can perfom both actions in an arbitrary order and it is not possible that performing one of these actions alone prevents $ \sigma^k\left( P \right) $ form performing the other one next.}, as depicted in Figure \ref{fig:confluence} process $ \sigma^k\left( P \right) $ must be able to perform both actions consecutively in arbitrary order resulting in the same term which we denote by $ Q_k $.
			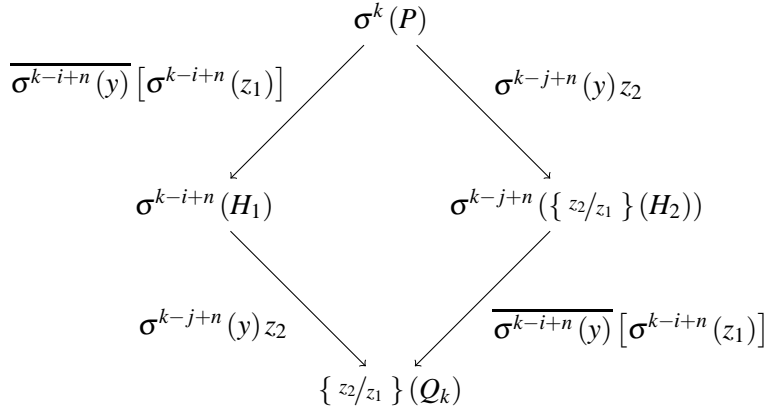
\begin{figure}[ht]
				\begin{center}
					\begin{tikzpicture}[node distance=3.5cm, auto]
						\node (a)						{$ \sigma^k\left( P \right) $};
						\node (b) [below left of=a]		{$ \sigma^{k{-}i{+}n}\left( H_1 \right) $};
						\node (c) [below right of=a]	{$ \sigma^{k{-}j{+}n}\left( \Set{ \Subs{z_2}{z_1} }\left( H_2 \right) \right) $};
						\node (d) [below right of=b]	{$ \Set{ \Subs{z_2}{z_1} }\left( Q_k \right) $};
						
						\path[->] (a) edge [swap]	node {$ \outP{\sigma^{k{-}i{+}n}\left( y \right)}{\sigma^{k{-}i{+}n}\left( z_1 \right)} $}		(b);
						\path[->] (a) edge			node {$ \sigma^{k{-}j{+}n}\left( y  \right) z_2 $}	(c);
						\path[->] (b) edge [swap]	node {$ \sigma^{k{-}j{+}n}\left( y  \right) z_2 $}	(d);
						\path[->] (c) edge			node {$ \outP{\sigma^{k{-}i{+}n}\left( y \right)}{\sigma^{k{-}i{+}n}\left( z_1 \right)} $}		(d);
					\end{tikzpicture}
				\end{center}
				\caption{Local confluence of receiving and sending actions.} \label{fig:confluence}
			\end{figure}
			To restore symmetry we build a chain of $ n $ steps such that each process $ \sigma^k\left( P \right) $ performs the output action $ \mu_{\mathsf{out}, k} $ in step $ \left( \left( k{-}i{+}n \right) \!\!\! \mod n \right){+}1 $ and the input action $ \mu_{\mathsf{in}, k} $ in step $ \left( \left( k{-}j{+}n \right) \!\!\! \mod n \right){+}1 $, i.e., each process is once a sender and once a receiver and $ \mu = \mu_2 = \ldots = \mu_n = \tau $ and with that $ \mu, \mu_2, \ldots, \mu_n = \SymSequence{\mu}{\sigma}{n, \tilde{x}} $. Again we consider the case of unbound outputs by first. Then we have $ \SymNetwork{P}{\sigma}{n, \tilde{x}} \step{\tau} \Set{ \Subs{H_1}{\sigma^i\left( P \right)}, \Subs{H_2}{\sigma^j\left( P \right)} }\SymNetwork{P}{\sigma}{n, \tilde{x}} $ as first step with $ \sigma^i\left( P \right) \step{\outU{y}{z_1}} H_1 $, $ \sigma^j\left( P \right) \step{y z_2} \Set{ \Subs{z_2}{z_1} }\left( H_2 \right) $ and $ \sigma^i\left( P \right) \mid \sigma^j\left( P \right) \step{\tau} H_1 \mid H_2 $. Depending on the values of $ i $ and $ j $ some of the processes perform the corresponding input action first while others perform at first the corresponding output action. Because of Lemma \ref{lem:localConfluence} both is possible. We let each process perform exactly these two actions (compare to Figure \ref{fig:confluence}). We choose $ P' = Q_0 $ and $ \tilde{x}' = \tilde{x} $. We start with a symmetric network and all processes behave symmetricaly, i.e., each process mimic the behavoir of its neighbour, so we have $ Q_k = \sigma^k\left( Q_0 \right) $ for all $ k $ with $ 0 \leq k \leq n{-}1 $ such that can choose $ \sigma' = \sigma $ and have
			\begin{align*}
				\SymNetwork{P}{\sigma}{n, \tilde{x}} \step{\tau}^n \SymNetwork{P'}{\sigma'}{n, \tilde{x}'} \text{.}
			\end{align*}
			
			Now we consider the case of bound outputs. Note that $ \sigma^i\left( P \right) $ and $ \sigma^j\left( P \right) $ peform a communication step \emph{within} the network, so if $ \sigma^i\left( P \right) $ performs a bound output $ z_1 $ must be bound in $ \sigma^i\left( P \right) $, i.e., $ z_1 \notin \tilde{x} $. By symmetry $ \sigma^l\left( z_1 \right) = z_1 $ is bound in $ \sigma^{i{+}l}\left( P \right) $ for all $ l \in \Set{ 0, \ldots, n{-}1 } $. With that either all output action are bound or all are unbound. In case of bound output we have $ \sigma^i\left( P \right) \mid \sigma^j\left( P \right) \step{\tau} \newVar{z, z'}{\left( H_1 \mid H_2 \right)} $, because first we have to apply alpha-conversion to rename the instance of $ z_1 $ bound in $ \sigma^j\left( P \right) $ and then the bound output by $ \sigma^i\left( P \right) $ leads to a scope extrusion such that $ z = z_1 $ and $ z' $ is the renaming of $ z_1 $ in $ \sigma^j\left( P \right) $. Again we  use alpha-conversion after the first communication step to rename all instances of $ z_1 $ (except the first) to a different fresh name for each process of the network and add the respective permutations of $ z_1 $ to $ \sigma $ in order to obtain $ \sigma' $ such that $ \sigma \subseteq \sigma' $. Let $ z_{1, 2}, \ldots, z_{1, n} $ denote the sequence of names used to rename $ z_1 $ according to $ \sigma' $. We proceed as in the case of unbound outputs with the $ n {-} 1 $ communication steps as described above. Of course we have to replace the processes $ \sigma^k\left( P \right) $ by $ \Set{ \Subs{z_{1, 2}}{z_1}, \ldots, \Subs{z_1}{z_{1, n}} }^{k{-}i}\left( P \right) $ and $ \mu_{\mathsf{out}, k} $ by $ \outP{\sigma^{k{-}i{+}n}\left( y \right)}{\Set{ \Subs{z_{1, 2}}{z_1}, \ldots, \Subs{z_1}{z_{1, n}} }^{k{-}i}\left( z_1 \right)} $ for $ 0 \leq k \leq n{-}1 $. After completing these $ n $ communication steps the names $ z_1, z_{1, 2}, \ldots, z_{1, n} $ are pulled outwards by scope extrusion, i.e. we have $ \SymNetwork{P}{\sigma}{n, \tilde{x}} \step{\tau} \Set{ \HOSubs{H_1}{i}, \HOSubs{H_2}{j} }\SymNetwork{P}{\sigma}{n, \tilde{x}, z_1, z_{1, 2}} \step{\tau}^{n{-}1} \SymNetwork{R}{\sigma'}{n, \tilde{x}'} $, where $ \tilde{x}' = \tilde{x}, z_1, z_{1, 2}, \ldots, z_{1, n} $ and $ P \step{\outB{\sigma^{n{-}i}\left( y \right)}{z_1}, \sigma^{n{-}j}\left( y \right)z_2} R $. With that we can choose $ P' = R $ and are done.
	\end{description}
\end{proof}

	With Lemma \ref{lem:cannotBreakSymmetry}, we can now construct the symmetric execution. We start with an arbitrary symmetric network $ \SymNetwork{P}{\sigma}{n, \tilde{x}} $. If $ \SymNetwork{P}{\sigma}{n, \tilde{x}} \not\step{} $ we have a symmetric execution of length $ 0 $. Otherwise, if $ \SymNetwork{P}{\sigma}{n, \tilde{x}} $ can perform a step labeled by $ \mu_1 $ by Lemma \ref{lem:cannotBreakSymmetry} we can perform $ n {-} 1 $ more steps such that $ \SymNetwork{P}{\sigma}{n, \tilde{x}} \step{\SymSequence{\mu_1}{\sigma_1}{n, \tilde{x}}} \SymNetwork{P_1}{\sigma_1}{n, \tilde{x}_1} $. Now we can proceed alike with $ \SymNetwork{P_1}{\sigma_1}{n, \tilde{x}_1} $ and result either in a finite symmetric execution of length $ 1 $ or we have $ \SymNetwork{P}{\sigma}{n, \tilde{x}} \step{\SymSequence{\mu_1}{\sigma_1}{n, \tilde{x}}} \SymNetwork{P_1}{\sigma_1}{n, \tilde{x}_1} \step{\SymSequence{\mu_2}{\sigma_2}{n, \tilde{x}_1}} \SymNetwork{P_2}{\sigma_2}{n, \tilde{x}_2} $. By recursively repeating this argument, we either get a finite symmetric execution of length $ m $ or an infinite symmetric execution.
\end{proof}

\paragraph{Breaking Symmetries.}
\label{sec:breaking-symmetries}

Note that Theorem \ref{thm:cannotBreakSymmetry} does not state anything about encodability and it does not need a notion of reasonableness either. Instead, it just states without any precondition that every symmetric network in \procsep has at least one symmetric execution.  In contrast, there are symmetric networks in $ \procmix $ without such a symmetric execution, as the following example shows. Consider the network
\begin{align*}
	\newVar{x, y}{\left( P \mid \sigma\left( P \right) \right)} \quad \text{ with } \quad P = \out{x}.\out{1} + y.\out{2} \quad \text{ and } \quad \sigma = \Set{ \Subs{x}{y}, \Subs{y}{x}, \Subs{1}{2}, \Subs{2}{1} }
\end{align*}
with $ \sigma^2 = \id $, i.e., $ \newVar{x, y}{\left( P \mid \sigma\left( P \right) \right)} $ is a symmetric network in $ \procmix $. It has, modulo structural congruence, exactly the two following executions 
\begin{align*}
	\newVar{x, y}{\left( P \mid \sigma\left( P \right) \right)} & \step{\tau} \out{1} \mid \out{1} \step{\out{1}} \out{1} \step{\out{1}} \nullTerm\\
	\newVar{x, y}{\left( P \mid \sigma\left( P \right) \right)} & \step{\tau} \out{2} \mid \out{2} \step{\out{2}} \out{2} \step{\out{2}} \nullTerm
\end{align*}
and even none of them is symmetric; the initial symmetry is broken.  So Theorem \ref{thm:cannotBreakSymmetry} proves a difference in the absolute expressive power between \pisep and \pimix\footnote{Remember that \pisep is a subset of \pimix and with it \pimix is at least as expressive as \pisep.}.

\fact{fac:absolutExpressivness}{
	The full $ \pi $-calculus is strictly more expressive as the $ \pi $-calculus without mixed choice.
}

\section{Non-Existence of Uniform Encodings} \label{sec:nonExistUniformEncoding}

As done by Palamidessi \cite{palamidessi03} and also by Gorla \cite{gorla08d}, we now also prove that there is no uniform and reasonable encoding from \pimix into \pisep, but here using Theorem \ref{thm:cannotBreakSymmetry} which states a difference in the absolute expressive power of the two calculi.  It is no real surprise that this absolute result leads to differences in the translational expressiveness of the languages. Because uniform encodings preserve symmetries|or at least enough of the symmetric nature of the terms|, the non-existence of a uniform and reasonable encoding is a natural consequence of the difference in their absolute expressiveness. Unfortunately, there is no agreement on the minimal requirements of a reasonable encoding, so we can not formally prove this result in general, although we belive that it holds for any meaningful Definition of reasonableness. Instead to underpin our assertion we prove it in the settings of \cite{palamidessi03} and \cite{gorla08d}.

According to \cite{palamidessi03}, an encoding is uniform if it translates the parallel operator homomorphically and preserves renamings, i.e., for all permutations of names $ \sigma $ there exists a permutation of names $ \theta $ such that $ \encoding{\sigma\left( P \right)} = \theta\left( \encoding{P} \right) $. Vigliotti et al.~\cite{vigliottiPhillipsPalamidessi07} additionally require that the permutations $ \sigma $ and $ \theta $ are compatible on observables. Gorla \cite{gorla08d} does not use the notion of uniformity, but in his first setting the separation result between \pimix and \pisep does also assume homomorphical translation of the parallel operator. Moreover, he specifies name invariance as a criteriaon for a good encoding, which is a more complex condition than  Palamidessi's second condition. It turns out that, in our setting, we do not need a second condition like renaming preservation or name invariance, because we base our counterexamples in the following separation results on symmetric networks of the form $ P \mid P $ as already Gorla did in \cite{gorla08d}.  For us, an encoding is uniform iff it translates the parallel operator homomorphically.

\definition{def:uniformEncoding}{Uniform encoding}{
	An encoding $ \nEncoding $ from $ \pimix $ into an other language is a \textit{uniform encoding} if and only if for all $ P, Q \in \procmix $
	\begin{align*}
		& \encoding{P \mid Q} = \encoding{P} \mid \encoding{Q} \tag{U} \label{uniform1}
	\end{align*}
}

Actually, Theorem \ref{thm:cannotBreakSymmetry} should suffice to prove that there can not be a uniform and reasonable encoding from \pimix into \pisep, because uniform encodings preserve symmetries and it is possible to break symmetries in \pimix while this is not possible in \pisep. The crux is that there is no commonly accepted notion of reasonableness. For separation results, we seek a definition of reasonableness that is as weak as possible. But, without any notion of reasonableness, the theorem would not hold, because there are uniform encodings from \pimix into \pisep. For instance, we could simply translate everything to $ \nullTerm $. Of course such an encoding makes no sense and so hardly anyone would call it reasonable. Usually, an encoding is called reasonable if it preserves some kind of behavior or the ability to solve some kind of problem so to ensure that the purpose of the original term is preserved.  In the following, we consider three different notions of reasonableness.

\paragraph{Version 1}

For Palamidessi, an encoding is reasonable if it preserves the relevant observables and termination properties \cite{palamidessi03}. Implicitly, she requires that a reasonable encoding should at least preserve the ability to solve leader election. We  do alike but with a different interpretation of what it means to solve leader election \ifadraft\footnote{\textcolor{red}{Fussnote nur einfügen, wenn die Aussage in einem TechRep gezeigt wird?!} If we would like to prove the following separation result with the Definition of leader election of \cite{palamidessi03} we would need to strenghten our Definition of uniformity. More precisely we would have to require that a uniform encoding is name invariant as defined by Gorla in \cite{gorla08d}. Name invariance allows us to prove that the uniform encoding of an arbitrary symmetric network $ \SymNetwork{P}{\sigma}{n, \tilde{x}} $ is again a symmetric network $ \SymNetwork{P}{\sigma'}{n, \tilde{y}} $ such that $ \sigma\left( i \right) = i $ iff $ \sigma'\left( \varphi\left( i \right) \right) = \varphi\left( i \right) $ for all names $ i $, where $ \varphi $ is a renaming policy as defined in \cite{gorla08d}. With that we can prove that each encoding of a symmetric network as e.g. $ \out{a}.\outU{\mathit{out}}{\; 1} + b.\outU{\mathit{out}}{\; 2} \mid \out{b}.\outU{\mathit{out}}{\; 2} + a.\outU{\mathit{out}}{\; 1} $ has at least one execution in which a decision on a leader $ \outU{\mathit{out}}{\; i} $ is directly followed by $ \outU{\mathit{out}}{\; \sigma'\left( i \right)} $ with $ \sigma'\left( i \right) \neq i $, contradicting leader election. Hence there is no uniform and reasonable encoding from \pimix into \pisep.}\fi that is more closely related to the definition used by Boug\'{e} \cite{bouge88}:
A network is said to solve leader election iff in each execution exactly one process propagates itself as leader while all the other processes propagate themselves as slaves. We assume the existence of two different predetermined output actions, one to propagate as leader and the other to propagate as slave. Moreover, we require that for both output actions neither the channel names nor the sent values are bound within the network\footnote{Note that if we allow bound names in these output actions, we could hardly predetermine them.}. The main difference to the definition of leader election used in \cite{palamidessi03} is that here the slaves do not have to know the identity, i.e., the index, of the leader. So, this definition is usually considered as a weaker notion of the leader election problem.  An encoding is now said to be reasonable iff it preserves the ability to solve the leader election problem.

\definition{def:Reasonableness1}{1-Reasonableness}{
	An encoding $ \nEncoding :\procmix\to\procsep$ is \textit{1-reasonable}, if $ \encoding{P} $ solves leader election if and only if $ P $ solves leader election for all $ P \in \procmix $.
}

To prove that there is no uniform and reasonable encoding we force our encoding to lead to a network of two processes that is symmetric with respect to identity. By Theorem \ref{thm:cannotBreakSymmetry}, this network has at least one symmetric execution. Because we use the identity as symmetry relation, in the symmetric execution both processes behave exactly the same such that if one of them propagates himself as leader then the other one does alike, which contradicts leader election.

\theorem{Separation Result}{thm:noUniformEncoding1}{
	There is no uniform and 1-reasonable encoding from \pimix into \pisep.
}
\begin{proof}[Proof of Theorem \ref{thm:noUniformEncoding1}]
	Let us assume the contrary, i.e., there is a uniform and 1-reasonable encoding $ \nEncoding $ from \pimix into \pisep. Consider the network:
	\begin{align*}
		N \; \triangleq \; P \mid P \quad \text{ with } \quad P \; \triangleq \; a.\out{\mathit{slave}} + \out{a}.\out{\mathit{leader}}
	\end{align*}
	Obviously $ \sigma = \id $ is a symmetry relation of degree $ 2 $ and so $ N = \SymNetwork{a.\out{\mathit{slave}} + \out{a}.\out{\mathit{leader}}}{\sigma}{2} $ is a symmetric network. Moreover $ N $ solves leader election, because the leader sends an empty message over channel $ \mathit{leader} $ and all slaves send an empty message over channel $ \mathit{slave} $. By Definition \ref{def:uniformEncoding} of uniformity, we have $ \encoding{P \mid P} \stackrel{(\ref{uniform1})}= \encoding{P} \mid \encoding{P} = \SymNetwork{\encoding{P}}{\id}{2} $, i.e., $ \encoding{N} $ is again a symmetric network of degree $ 2 $ with $ \id $ as symmetry relation. By Theorem \ref{thm:cannotBreakSymmetry}, $ \encoding{N} $ has at least one symmetric execution and by reasonableness $ \encoding{N} $ must solve leader election, i.e., there is exactly one process that propagates itself as leader by an output action. Let $ \mu_l $ denote this send action. By Definition \ref{def:symmetricTrace}, a symmetric execution has symmetric sequences of actions, i.e., the action $ \mu_l $ is coupled to its symmetric counterpart building the sequence $ \SymSequence{\mu_l}{\sigma'}{2, \tilde{z}'} $ for some $ \tilde{z}' \in \tupel{\names} $ and $ \sigma' \in \symRels{2}{\names} $. By construction in the proof of Lemma \ref{lem:cannotBreakSymmetry}, and because we start with $ \id $, we know that $ \sigma' $ consists of (permutations of) names that are bound in $ \encoding{N} $ or fresh. Because, by definition, $ \mu_l $ can neither contain fresh nor bound names, we conclude $ \SymSequence{\mu_l}{\sigma'}{2, \tilde{z}'} = \mu_l, \mu_l $, i.e., the output action appears twice in the symmetric execution. With that two processes propagate themselves as leader, which is a contradiction.
\end{proof}

Note that, in contrast to the proof of Palamidessi \cite{palamidessi03,vigliottiPhillipsPalamidessi07}, we do not have to assume that the encoding is renaming preserving.% and \textcolor{red}{reflects divergence. Of course this significantly strengthens the statement of the proof.} \K{Stimmt das? Ich verstehe immer noch nicht wozu Palamidessi Divergence-Freiheit braucht. Das sollten wir klären.}

\paragraph{Version 2}

Here, we first introduce a technical lemma. Intuitively, it states that the symmetric execution of a symmetric network of degree $ n $, where $ n $ is \emph{not} the minimal degree of the corresponding symmetry relation, can be subdivided into symmetric executions on symmetric subnetworks of the original network.
\lemma{lem:subdividingSymmetricTraces}{
	Let $ \SymNetwork{P_0}{\sigma}{n, \tilde{x}} $ be a symmetric network in \procsep. If the degree of $ \sigma $ is not minimal, i.e., if there is a $ n' \in \Nat $ with $ 0 < n' < n $ such that $ \sigma^{n'} = \id $, then $ \SymNetwork{P_0}{\sigma}{n, \tilde{x}} $ has a finite or an infinite symmetric execution
	\begin{align*}
		\SymNetwork{P_0}{\sigma}{n, \tilde{x}} \step{\SymSequence{\mu_1}{\sigma_1}{n, \tilde{x}}} \SymNetwork{P_1}{\sigma_1}{n, \tilde{x}_1} \step{\SymSequence{\mu_2}{\sigma_2}{n, \tilde{x}_1}} \ldots \step{\SymSequence{\mu_m}{\sigma_m}{n, \tilde{x}_{m{-}1}}} \SymNetwork{P_m}{\sigma_m}{n, \tilde{x}_m} \not\step{} \quad \OR \quad \SymNetwork{P_0}{\sigma}{n, \tilde{x}} \step{\SymSequence{\mu_1}{\sigma_1}{n, \tilde{x}}} \SymNetwork{P_1}{\sigma_1}{n, \tilde{x}_1} \step{\SymSequence{\mu_2}{\sigma_2}{n, \tilde{x}_1}} \ldots
	\end{align*}
	for a $ m \in \Nat $, $ P_1, \ldots, P_m \in \procsep $, $ \sigma_1, \ldots, \sigma_m \in \symRels{n}{\names} $ with $ \sigma \subseteq \sigma_1 \subseteq \ldots \subseteq \sigma_m $, $ \tilde{x}_1, \ldots, \tilde{x}_m \in \tupel{\names} $ and $ \mu_1, \ldots, \mu_m \in \lab $ or some $ P_1, P_2, \ldots \in \procsep $, $ \sigma_1, \sigma_2, \ldots \in \symRels{n}{\names} $ with $ \sigma \subseteq \sigma_1 \subseteq \sigma_2 \subseteq \ldots $, some $ \tilde{x}_1, \tilde{x}_2, \ldots \in \tupel{\names} $ and $ \mu_1, \mu_2, \ldots \in \lab $ respectively such that $ \SymNetwork{P_0}{\sigma}{n', \tilde{x}} $ has the finite or infinite symmetric execution
	\begin{align*}
		\SymNetwork{P_0}{\sigma}{n', \tilde{x}'} \step{\SymSequence{\mu_1'}{\sigma_1'}{n', \tilde{x}'}} \SymNetwork{P_1}{\sigma_1'}{n', \tilde{x}_1'} \step{\SymSequence{\mu_2'}{\sigma_2'}{n', \tilde{x}_1'}} \ldots \step{\SymSequence{\mu_m'}{\sigma_m'}{n', \tilde{x}_{m{-}1}'}} \SymNetwork{P_m}{\sigma_m'}{n', \tilde{x}_m'} \not\step{} \quad \OR \quad \SymNetwork{P_0}{\sigma}{n', \tilde{x}} \step{\SymSequence{\mu_1'}{\sigma_1'}{n', \tilde{x}}} \SymNetwork{P_1}{\sigma_1'}{n', \tilde{x}_1'} \step{\SymSequence{\mu_2'}{\sigma_2'}{n', \tilde{x}_1'}} \ldots
	\end{align*}
	for some $ \tilde{x}_1', \ldots, \tilde{x}_m' \in \tupel{\names} $, $ \mu_1', \ldots, \mu_m' \in \lab $ and $ \sigma_1', \ldots, \sigma_m' \in \symRels{n'}{\names} $ with $ \sigma \subseteq \sigma_1' \subseteq \ldots \subseteq \sigma_m' $ or some $ \tilde{x}_1', \tilde{x}_2', \ldots \in \tupel{\names} $, $ \mu_1', \mu_2', \ldots \in \lab $ and $ \sigma_1', \sigma_2', \ldots \in \symRels{n'}{\names} $ with $ \sigma \subseteq \sigma_1' \subseteq \sigma_2' \subseteq \ldots $ respectively such that $ \tilde{x}' $ is a subsequence of $ \tilde{x} $, $ \tilde{x}_i' $ is a subsequence of $ \tilde{x}_i $ and either $ \mu_i' $ or if $ \mu_i' $ is a bound output its unbound variant is in $ \SymSequence{\mu_i}{\sigma_i}{n, \tilde{x}_{i{-}1}} $ for all $ i \in \Set{ 1, \ldots, m } $ or $ i \in \Nat $ respectively.
}
\noindent 
Note that, like Theorem \ref{thm:cannotBreakSymmetry}, this result is absolute in the sense that it holds independently of any notion of uniformity or reasonableness. 
%We only present a proof sketch here.  A proof can be found in the appendix.
%
%\begin{proof}[Proof Sketch of Lemma \ref{lem:subdividingSymmetricTraces}]
%	Assume there is a $ 0 < n' < n $ such that $ \sigma^{n'} = \id $. Then because $ \sigma^n = \id $ there must be a $ k \in \Nat $ such that $ n = k * n' $. Because $ \sigma^0 = \sigma^{n'} = \sigma^{i * n'} $ for each $ i \in \Set{ 1, \ldots, k } $ we have $ \sigma^j = \sigma^{j{+}n'} $. So $ \SymNetwork{P_0}{\sigma}{n, \tilde{x}} $ can be divided into $ k $ identical symmetric networks such that $ \SymNetwork{P_0}{\sigma}{n, \tilde{x}} = \SymNetwork{P_0}{\sigma}{n', \tilde{x}} \mid \ldots \mid \SymNetwork{P_0}{\sigma}{n', \tilde{x}} $ and $ \SymSequence{\mu_1}{\sigma_1}{n, \tilde{x}} $ can be divided into $ k $ identical sequences $ \SymSequence{\mu_1'}{\sigma_1'}{n', \tilde{x}'} $ for each $ P_0 \in \procsep $ and each $ \mu_1 \in \lab $.
%	
%	The proof is by induction on the number of sequences of $ n $ steps from a symmetric network to a symmetric network. To prove the inductive step, we perform a case analysis on whether the first step of such a sequence is due to an action of only one process of the network or to a communication between two processes.
%\end{proof}

\begin{proof}[Proof of Lemma \ref{lem:subdividingSymmetricTraces}]
	Assume there is a $ 0 < n' < n $ such that $ \sigma^{n'} = \id $. Then because $ \sigma^n = \id $ there must be a $ k \in \Nat $ such that $ n = k * n' $. Because $ \sigma^0 = \sigma^{n'} = \sigma^{i * n'} $ for each $ i \in \Set{ 1, \ldots, k } $ we have $ \sigma^j = \sigma^{j{+}n'} $. So $ \SymNetwork{P'}{\sigma}{n, \tilde{x}} $ can be divided into $ k $ identical symmetric networks such that $ \SymNetwork{P'}{\sigma}{n, \tilde{x}} = \SymNetwork{P}{\sigma}{n', \tilde{x}} \mid \ldots \mid \SymNetwork{P}{\sigma}{n', \tilde{x}} $ and $ \SymSequence{\mu'}{\sigma}{n, \tilde{x}} $ can be divided in $ k $ identical sequences such that $ \SymSequence{\mu'}{\sigma}{n, \tilde{x}} = \SymSequence{\mu'}{\sigma}{n', \tilde{x}}, \ldots, \SymSequence{\mu'}{\sigma}{n', \tilde{x}} $ for each $ P' \in \procsep $ and each $ \mu' \in \lab $.
	
	If $ \SymNetwork{P_0}{\sigma}{n, \tilde{x}} $ has a symmetric execution of length $ 0 $, i.e., $ \SymNetwork{P_0}{\sigma}{n, \tilde{x}} \not \step{} $, then of course we have $ \SymNetwork{P_0}{\sigma}{n', \tilde{x}} \not \step{} $ as well and so $ \SymNetwork{P_0}{\sigma}{n', \tilde{x}} $ has a symmetric execution of length $ 0 $.
	
	Else we consider an arbitrary sequence of $ n $ steps $ \SymNetwork{P_k}{\sigma_k}{n, \tilde{x}_k} \step{\SymSequence{\mu_{k{+}1}}{\sigma_{k{+}1}}{n, \tilde{x}_k}} \SymNetwork{P_{k{+}1}}{\sigma_{k{+}1}}{n, \tilde{x}_{k{+}1}} $ of the given symmetric execution for $ k \in \Set{0, \ldots, m} $ in the case of a finite symmetric execution and $ k \in \Nat $ for an infinite symmetric execution. As constructed in Theorem \ref{thm:cannotBreakSymmetry} $ P_{k{+}1} $ is either the result of a step of $ \sigma_k^i\left( P_k \right) $ realized without the rules \textsc{Comm} and \textsc{Close} or it is the result of two communications of $ \sigma_k^i\left( P_k \right) $ and $ \sigma_k^j\left( P_k \right) $ realized by one of the rules \textsc{Comm} or \textsc{Close}. We proceed with a cases split.
	\begin{description}
		\item[Case without \textsc{Comm} and \textsc{Close}:] Let $ \sigma_k^i\left( P_k \right) $ with $ i \in \Set{ 0, \ldots, n{-}1 } $ be the process which performs the first of the $ n $ steps labeled $ \mu_{k{+}1} $. We choose $ \mu_{k{+}1}' $ as the $ n{-}i $'th action in $ \SymSequence{\mu_{k{+}1}}{\sigma_{k{+}1}}{n, \tilde{x}_k} $, i.e., we choose the label of the action performed by process $ P_k $. If $ \mu_{k{+}1} $ is a bound output and $ \mu_{k{+}1}' $ is not then we choose the bound ouput variant of $\mu_{k{+}1}' $. By construction in the proof of Lemma \ref{lem:cannotBreakSymmetry} there are $ n' $ steps performed by the processes $ \sigma_k^0\left( P_k \right) $, \ldots, $ \sigma_k^{n'{-}1}\left( P_k \right) $ and labeled by the first $ n' $ labels of $ \SymSequence{\mu_{k{+}1}'}{\sigma_{k{+}1}}{n, \tilde{x}_k} $. Note that because $ \sigma_k' $ differs from $ \sigma_k $ only on permutations on formerly bound names we can perform these steps by $ {\sigma_k'}^0\left( P_k \right) $, \ldots, $ {\sigma_k'}^{n'{-}1}\left( P_k \right) $, too. If $ \mu_k $ is no bound output we can choose $ \tilde{x}_{k{+}1}' = \tilde{x}_k' $ and $ \sigma_{k{+}1}' = \sigma_k' $ and are done. Else if $ \mu_k = \outU{y}{z} $ and $ z \notin \boundnames{\sigma_k^i\left( P_k \right)} $ we can choose $ \sigma_{k{+}1}' = \sigma_k' $ and $ \tilde{x}_{k{+}1}' $ as the sequence of names in $ \tilde{x}_k', z_1, \ldots, z_l $, where $ z_1, \ldots, z_l $ are the values of the bound outputs in $ \SymSequence{\mu_{k{+}1}'}{\sigma_{k{+}1}'}{n', \tilde{x}_k} $. Else if $ z \in \boundnames{\sigma_k^i\left( P_k \right)} $ we can choose $ \tilde{x}_{k{+}1}' = \tilde{x}_k' $ and we add the permutations of $ z $ done by alpha conversion as descriebed in Lemma \ref{lem:cannotBreakSymmetry} to $ \sigma_k' $ to obtain $ \sigma_{k{+}1}' $. Again by construction in Lemma \ref{lem:cannotBreakSymmetry} performing these $ n' $ steps steps we have $ \SymNetwork{P_k}{\sigma_k'}{n', \tilde{x}_k'} \step{\SymSequence{\mu_{k{+}1}'}{\sigma_{k{+}1}'}{n', \tilde{x}_k'}} \SymNetwork{P_{k{+}1}}{\sigma_{k{+}1}'}{n', \tilde{x}_{k{+}1}'} $.
		\item[Case with \textsc{Comm} or \textsc{Close}:] Then $ \SymSequence{\mu_{k{+}1}}{\sigma_{k{+}1}}{n, \tilde{x}_k} $ is a sequence of $ n $ times $ \tau $. We choose $ \mu_{k{+}1}' = \mu_{k{+}1} = \tau $ and $ \SymSequence{\mu_{k{+}1}'}{\sigma_k}{n', \tilde{x}_k} $ is a sequence of $ n' $ times $ \tau $. Let $ \sigma_k^i\left( P_k \right) $ and $ \sigma_k^j\left( P_k \right) $ with $ i, j \in \Set{ 0, \ldots, n{-}1} $ be the processes which perform the first of the $ n $ steps. Without lost of generality let $ \sigma_k^i\left( P_k \right) $ be the sender and $ \sigma_k^j\left( P_k \right) $ be the receiver, i.e., $ \sigma_k^i\left( P_k \right) $ performs an output action $ \out{\gamma} $ and $ \sigma_k^j\left( P_k \right) $ performs the complemtary receiving action $ \gamma $. By construction in the proof of therorem \ref{thm:cannotBreakSymmetry} the first $ n' $ steps within $ \SymNetwork{P_k}{\sigma_k}{n, \tilde{x}_k} \step{\SymSequence{\mu_{k{+}1}}{\sigma_{k{+}1}}{n, \tilde{x}_k}} \SymNetwork{P_{k{+}1}}{\sigma_{k{+}1}}{n, \tilde{x}_{k{+}1}} $ are performed by the senders $ \sigma_k^i\left( P_k \right) $, \ldots, $ \sigma_k^{i{+}n'{-}1}\left( P_k \right) $ in this order sending $ \SymSequence{\overline{\gamma}}{\sigma_{k{+}1}}{n', \tilde{x}_k} $ respectively and by the receivers $ \sigma_k^j\left( P_k \right) $, \ldots, $ \sigma_k^{j{+}n'{-}1}\left( P_k \right) $ in this order receiving $ \SymSequence{\gamma}{\sigma_{k{+}1}}{n', \tilde{x}_k} $. Now because of $ \sigma^{n'} = \id $ and $ \sigma_k' $ differs from $ \sigma $ only by formerly bound names and their renamings according to alph-conversion for each $ g \in \Set{ i, \ldots, i{+}n'{-}1 } $ and for each \linebreak $ h \in \Set{ j, \ldots, j{+}n'{-}1 } $ we have $ \sigma_k^g\left( P_k \right) $ and $ {\sigma_k'}^{g \!\!\! \mod n'}\left( P_k \right) $, and $ \sigma_k^h\left( P_k \right) $ and $ {\sigma_k'}^{h \!\!\! \mod n'}\left( P_k \right) $ respectively are equal modulo the renaming performed by formerly alpha-conversion. With that we can again close the cycle as in the proof of Lemma \ref{lem:cannotBreakSymmetry} leading to $ \SymNetwork{P_k}{\sigma_k}{n', \tilde{x}_k'} \step{\SymSequence{\mu_{k{+}1}'}{\sigma_{k{+}1}'}{n', \tilde{x}_k'}} \SymNetwork{P_{k{+}1}}{\sigma_{k{+}1}'}{n', \tilde{x}_{k{+}1}'} $, where $ \title{x}_{k{+}1}' $ and $ \sigma_{k{+}1}' $ are obtained from $ \title{x}_k' $ and $ \sigma_k' $ as descriebed in Lemma \ref{lem:cannotBreakSymmetry}.
	\end{description}
	Because we can subdivide an arbitrary sequence of $ n $ steps we can subdivide each such sequence in the symmetric execution and with it the symmetric execution.
\end{proof}

Gorla \cite{gorla08d} defines the reasonableness of an encoding by the properties operational correspondence, divergence reflection and success sensitiveness. We use just the last of his properties instantiated with must testing. So we implicitly require divergence reflection. According to \cite{gorla08d}, success is represented by a process $ \sqrt{} $ that is part of the source and the target language of the encoding and always appears unbound. More precisely, a process must-succeeds if it \emph{always} reduces to a proccess containing a top-level unguarded occurence of $ \sqrt{} $. The fact that $ P $ must-succeeds is denoted by $ P \downdownarrows $. With it, an encoding is reasonable if the encoding of a term must-succeeds iff the term itself must-succeeds.

\definition{def:Reasonableness2}{2-Reasonableness}{
	An encoding $ \nEncoding :\procmix\to\procsep $ is \textit{2-reasonable}, if $ P \downdownarrows $ iff $ \encoding{P} \downdownarrows $ for all $ P \in \procmix $.
}

Again, we choose a term such that the encoding results in a network of the form $ Q \mid Q $ in \procsep that is symmetric with respect to the identity. In this case, we take advantage of the fact that the minimal degree of $ \id $ is less than the degree of the network such that we can use Lemma \ref{lem:subdividingSymmetricTraces} to subdivide the symmetric execution. With it already $ Q $ can perform the same sequence of steps as each process in $ Q \mid Q $ performs in the symmetric execution.

\theorem{Separation Result}{thm:noUniformEncoding2}{
	There is no uniform and 2-reasonable encoding from \pimix into \pisep.
}
\begin{proof}[Proof of Theorem \ref{thm:noUniformEncoding2}]
	Let us assume the contrary, i.e., there is a uniform and 2-reasonable encoding $ \nEncoding $ from \pimix into \pisep. Consider the network:
	\begin{align*}
		N \; \triangleq \; P \mid P \quad \text{ with } \quad P \; \triangleq \; a.\nullTerm + \out{a}.\sqrt{}
	\end{align*}
	Obviously, $ \sigma = \id $ is a symmetry relation of degree $ 2 $ and so $ N = \SymNetwork{a.\nullTerm + \out{a}.\sqrt{}}{\sigma}{2} $ is a symmetric network. Moreover, we have $ N \downdownarrows $ but $ P \not\downdownarrows $. We have $ \encoding{P \mid P} \stackrel{(\ref{uniform1})}= \encoding{P} \mid \encoding{P} = \SymNetwork{\encoding{P}}{\id}{2} $, i.e., $ \encoding{N} $ is again a symmetric network of degree $ 2 $ with $ \id $ as symmetry relation. By Theorem \ref{thm:cannotBreakSymmetry}, $ \encoding{N} $ has at least one symmetric execution and by success sensitiveness and must testing $ \encoding{N} $ must reduce to a proccess containing a top-level unguarded occurence of $ \sqrt{} $ within this symmetric execution, i.e., there is a sequence of actions $ \tilde{\mu} \in \tupel{\lab} $, a process $ P' \in \procsep $, a $ \sigma' \in \symRels{2}{\names} $ and a sequence of names $ \tilde{x'} $ such that $ \encoding{P} \mid \encoding{P} \step{\tilde{\mu}} \SymNetwork{P'}{\sigma'}{2, \tilde{x}} $ and $ P' $ or $ \sigma'\left( P' \right) $ contain a top-level unguarded occurence of $ \sqrt{} $. Then, by symmetry, both processes of $ \SymNetwork{P'}{\sigma'}{2, \tilde{x}} $ contain a top-level unguarded occurence of $ \sqrt{} $. By Lemma \ref{lem:subdividingSymmetricTraces}, there is a sequence of actions $ \tilde{\mu}' \in \tupel{\lab} $ and an execution $ \encoding{P} \step{\tilde{\mu}'} \newVar{\tilde{x}'}{P'} $ for a subsequence $ \tilde{x}' $ of $ \tilde{x} $. With it, $ \encoding{P} \downdownarrows $, and with success sensitiveness $ P \downdownarrows $, which is a contradiction.
\end{proof}

Note that, reconsidering the proofs of this separation result in \cite{gorla08d}, we managed to omit one of Gorla's additional assumptions\footnote{Namely, we do not need the assumption that $ \asymp_2 $ is exact (first setting in \cite{gorla08d}) or reduction sensitive (second setting in \cite{gorla08d}) and we do not need to assume the stronger version of operational correspondence in the third setting in \cite{gorla08d}. On the other side Gorla does not need to assume homomorphical translation of $ | $ in his second and third setting. He uses the weaker notion of compositional translation of $ | $ instead.}. Moreover, note that because we focus on breaking symmetries instead of leader election, we can apply Theorem \ref{thm:cannotBreakSymmetry} to problem instances different from leader election.

\paragraph{Version 3}

In his proofs of this separation result in \cite{gorla08d} Gorla uses may testing to show that there are terms $ P \in \procmix $ such that $ P \not\step{} $, $ P \not\downdownarrows $ and $ \left( P \mid P \right) \downdownarrows $, but there are no such terms in \procsep. Implicitly, he uses the fact that $ P \not\downdownarrows $ and $ \left( P \mid P \right) \downdownarrows $ implies $ P \mid P \step{} $ and that there are no terms $ P $ in \procsep such that $ P \not\step{} $ and $ P \mid P \step{} $. By proving this fact directly, we do not need any notion of testing to prove the separation result.

\definition{def:Reasonableness3}{3-Reasonableness}{
	An encoding $ \nEncoding  :\procmix\to\procsep $ is \textit{3-reasonable} if $ P \step{} $ if and only if $ \encoding{P} \step{} $ for all $ P \in \procmix $.
}

As far as we know, only few intuitively reasonable encodings are not also 3-reasonable.

Again, for the separation proof, we enforce that the encoding results in a symmetric network $ Q \mid Q $. By subdividing the symmetric execution of this network, we prove that $ Q \step{} $ iff $ Q \mid Q \step{} $, which does not necessarily hold in \pimix.

\theorem{Separation Result}{thm:noUniformEncoding3}{
	There is no uniform and 3-reasonable encoding from \pimix into \pisep.
}
\begin{proof}[Proof of Theorem \ref{thm:noUniformEncoding3}]
	Let us assume the contrary, i.e., there is a uniform and 3-reasonable encoding $ \nEncoding $ from \pimix into \pisep. Consider the network:
	\begin{align*}
		N \; \triangleq \; P \mid P \quad \text{ with } \quad P \; \triangleq \; a + \out{a}
	\end{align*}
	Obviously, $ \sigma = \id $ is a symmetry relation of degree $ 2 $ and so $ N = \SymNetwork{a + \out{a}}{\sigma}{2} $ is a symmetric network. Moreover, we have $ N \step{} $ but $ P \not\step{} $. We have $ \encoding{P \mid P} \stackrel{(\ref{uniform1})}= \encoding{P} \mid \encoding{P} = \SymNetwork{\encoding{P}}{\id}{2} $, i.e., $ \encoding{N} $ is again a symmetric network of degree $ 2 $ with $ \id $ as symmetry relation. By Theorem \ref{thm:cannotBreakSymmetry} $ \encoding{N} $ has at least one symmetric execution and by 3-reasonableness we have $ \encoding{P} \mid \encoding{P} \step{} $ and $ \encoding{P} \not\step{} $. By Lemma \ref{lem:cannotBreakSymmetry}, $ \encoding{P} \mid \encoding{P} \step{} $ implies that there is at least one step in the symmetric execution, i.e., there is a $ \mu \in \lab $, a process $ P' \in \pisep $, a $ \sigma' \in \symRels{2}{\names} $ and a sequence of names $ \tilde{x} \in \tupel{\names} $ such that $ \encoding{P} \mid \encoding{P} \step{\SymSequence{\mu}{\sigma'}{2}} \SymNetwork{P'}{\sigma'}{2, \tilde{x}} $. By Lemma \ref{lem:subdividingSymmetricTraces}, there is a execution $ \encoding{P} \step{\mu'} \newVar{\tilde{x}'}{P'} $ for a subsequence $ \tilde{x}' $ of $ \tilde{x} $, $ \mu' \in \SymSequence{\mu}{\sigma'}{2} $, which is a contradiction.
\end{proof}

%Gorla \cite{gorla08d} proves that this condition can be derived by the conditions of operational correspondence, divergence sensitiveness and some form of termination sensitiveness, if operational correspondence is formulated up-to a reduction-sensitive equivalence, e.g. identity, structural equivalence, strong bisimilarity or expansion preorder. So, in this case, there is no need to require divergence reflection.
Note that in opposite to both Palamidessi and Gorla we do not even assume divergence reflection.

% Moreover it turns out that the main reason for the non-existence of a uniform and reasonable encoding is that the uniformity of the encoding preserves the symmetric nature of the original process|or at least enough of its symmetric nature|and as proved in Theorem \ref{thm:cannotBreakSymmetry} it is not possible to break symmetries in \pisep which is possible in \pimix. This disparity in the absolute expressive power of \pimix and \pisep naturally leads to a difference in their translational expressive power. With that as consequence of Theorem \ref{thm:cannotBreakSymmetry} a reasonable encoding from \pimix into \pisep have to be able to break symmetries.

% \fact{fac:encodingBreakSymmery}{
% 	Any reasonable encoding from \pimix into \pisep must be able to break symmetries.
% }

\section{Conclusion and Future Work} \label{sec:conclusion}

We prove that \pimix is strictly more expressive than \pisep by means of an absolute separation result about the ability to break initial symmetries. This result is inpendent of any notion of encodibility, uniformity and reasonableness. By choosing the problem of breaking initial symmetries instead of leader election, we may significantly weaken the underlying definition of symmetry in comparison to \cite{palamidessi03}.  Moreover, we could still apply our absolute separation result to derive that there is no uniform and reasonable encoding from \pimix into \pisep considering three different definitions of reasonableness.  It turns out that the concentration on the underlying problem of breaking initial symmetries allows us to use counterexamples different from leader election to prove the translational separation results. Likewise, the separation result in the setting of \cite{gorla08d} can be derived by our absolute separation result as well. Besides that, our absolute separation result allows us to weaken the definition of uniformity in comparison to the translational separation result of \cite{palamidessi03}, and also to weaken the definition of reasonableness in comparison to the translational separation result in the first setting of \cite{gorla08d}.  Moreover, considering our last translational separation result, we can even withdraw the assumption of divergence reflection.

Our own translational separation results, i.e., the proofs of the non-existence of a uniform and reasonable encoding for different definitions of reasonableness, follow similar lines of argument. The proofs argue by contradiction. First, a symmetric network of the form $ P \mid P $ in \procmix with special features is presented.  Second, we use the fact that uniformity, i.e., the homomorphic translation of the parallel operator, preserves essentials parts of the symmetric nature of $ P \mid P $.  Third, we apply Theorem \ref{thm:cannotBreakSymmetry} to conclude with the existence of a symmetric execution. In two proofs, we then apply Lemma \ref{lem:subdividingSymmetricTraces} to subdivide this symmetric execution. At last, we derive a contradiction between the additional information provided by the symmetric execution (and its subdivision) and the respective definition of reasonableness.

Note that we prove the absolute result without any precondition.  We use different definitions of reasonableness for the translational results. The only constant precondition of the translational separation results is the definition of uniformity, i.e., the homomorphic translation of the parallel operator. This condition is crucial. Without it, we could not apply our absolute separation result. To the best of our knowledge, only Gorla ever managed to prove such a separation result between \pimix and \pisep without the homomorphic translation of the parallel operator, using compositionality, operational correspondence, divergence reflection, success sensitiveness and either a reduction sensitive version of $ \asymp $ or the stronger version of operational correspondence of his third setting.  However, Gorla believes that the result also holds for the general formulation of his criteria, i.e., without assuming a reduction sensitive version of $ \asymp $ or the stronger version of operational correspondence of his third setting. We believe that this is an interesting open question.

%\K{Das sollten wir erst aufgreifen, wenn wir eine Kodierung haben oder wissen, dass es keine gibt.} \textcolor{red}{Vielleicht wäre noch ein Text schön, der ein bisschen herraus arbeitet, wie genaus sich das absolute Resultat auf die translational Resultate auswirkt. Ich weiß von keinem Beweis zwischen \pimix und \pisep der nicht auf die eine oder andere Art Symmetrien und das Brechen von Symmetrien ausnutzt, auch wenn ich das nur schwer zeigen kann. Wenn die Fähigkeit Symmetrien zu brechen der einzige Unterschied zwischen \pimix und \pisep ist, dann muss es quasi nicht uniforme Kodierungen geben, es sei denn das man zeigen kann, dass es nicht möglich ist eine Kodierung anzugeben die alle Fälle von Symmetrien bricht ohne dabei das beobachtbare Verhalten des Terms zu ändern.}

% \textcolor{red}{Palamidessi argumentiert für die homomorphe Übersetzung des Parallel-Operators, weil diese den Grad der Verteiltheit bewahrt; an verschiedenen Stellen wurde bereits argumentiert, dass das vielleicht eine etwas zu harte Forderung ist; zumal auch andere Forderungen den Grad der Verteiltheit bewahren; Gorla argumentiert warum copositional auch gut ist und eine compositionale Kodierung verändert auch nicht zwangsweise den Grad der Verteiltheit; mit einer kompositionalen Kodierung würden die Separation Results aber nicht mehr durchgehen; Soll ich mich dazu auch noch auslassen?}

We may also turn the non-existence of a uniform \emph{and} reasonable encoding around and rephrase it as a weakened existence statement.  Recall that any uniform encoding from \pimix into \pisep preserves symmetries.  While it is possible to break such symmetries in \pimix, this is not possible in \pisep.  Thus, should there be a non-uniform (at least: ``weakly compositional'') \emph{but} reasonable encoding from \pimix into \pisep, then \emph{it would have to be the encoding itself to break these symmetries}.  Finding such a reasonable encoding is an open problem, if reasonableness includes divergence reflection.  A uniform and ``almost reasonable'' divergent encoding was already presented in \cite{nestmann00}.

\addcontentsline{toc}{section}{References}
\bibliographystyle{alpha}
\bibliography{BreakingSymmetries.bib}

\end{document}